\documentclass[aps,prd,twoside,twocolumn,nofootinbib,showpacs,floatfix,superscriptaddress]{revtex4-1}
\usepackage{amsfonts}
\usepackage{amssymb}
\usepackage{amsmath,amssymb}
\usepackage{graphicx,bm}
\usepackage{slashed}
\usepackage{times}
\usepackage{epstopdf}
\usepackage{ulem} 
\usepackage[usenames]{color}
\usepackage{float}
\usepackage{subfigure}
\usepackage{subfigure}
\usepackage{rotating}
\usepackage{color}
\usepackage{multirow}
\usepackage{dcolumn}
\usepackage{overpic}
\usepackage{booktabs}
\usepackage{makecell}
\usepackage{diagbox}

\renewcommand\sout{\bgroup \color{red} \ULdepth=-.5ex \ULset}

\usepackage[colorlinks, citecolor=blue,anchorcolor=red,menucolor=red,linkcolor=blue,filecolor=red,runcolor=red,urlcolor=blue,frenchlinks=red]{hyperref}
\allowdisplaybreaks[4]

\begin{document}
\title{$\Sigma_c\Sigma_c$ interactions in chiral effective field theory}
\author{Kan Chen}
\affiliation{School of Physics and Center of High Energy Physics,
Peking University, Beijing 100871, China}
\author{Bo-Lin Huang}
\affiliation{School of Physics and Center of High Energy Physics,
Peking University, Beijing 100871, China}
\author{Bo Wang}\email{wangbo@hbu.edu.cn}
\affiliation{School of Physical Science and Technology, Hebei
University, Baoding 071002, China\\
Key Laboratory of High-precision Computation and Application of
Quantum Field Theory of Hebei Province, Baoding 071002, China }
\author{Shi-Lin Zhu}\email{zhusl@pku.edu.cn}
\affiliation{School of Physics and Center of High Energy Physics,
Peking University, Beijing 100871, China}
\begin{abstract}
We study the interactions of the $\Sigma_c\Sigma_c$ system in the
framework of chiral effective theory. We consider the contact,
one-pion and two-pion exchange interactions and bridge the low
energy constants of the $\Sigma_c\Sigma_c$ system to those of the
$\Sigma_c^{(*)}\bar{D}^{(*)}$ systems through the quark-level ansatz
for the contact interaction. We explore the influence of
intermediate channels in the two-pion exchange diagrams of the
$\Sigma_c\Sigma_c$ system. We obtain a deep bound state
$[\Sigma_c\Sigma_c]_{J=0}^{I=0}$ and a shallow bound state
$[\Sigma_c\Sigma_c]_{J=1}^{I=1}$. As a byproduct, we further
investigate the interactions of the $\Lambda_c\Lambda_c$ and
$\Lambda_c\Sigma_c$ systems.
\end{abstract}
\maketitle
\section{Introduction}\label{sec1}

The quark model described the conventional mesons and baryons quite
well \cite{Godfrey:1985xj,Capstick:1985xss}. However, since the
discoveries of the $D_{s0}^*(2317)$ \cite{BaBar:2003oey} and
$X(3872)$ \cite{Belle:2003nnu} in 2003, a large number of states
that cannot be classified into conventional hadrons are continuously
reported~\cite{ParticleDataGroup:2020ssz}.  Different pictures have
been proposed to understand the nature of these exotic states, such
as the compact multiquark states, loosely bound hadronic molecules,
and kinetic effects,
etc~\cite{Chen:2016qju,Liu:2019zoy,Liu:2013waa,Guo:2017jvc,Hosaka:2016pey,Lebed:2016hpi,Esposito:2016noz,Brambilla:2019esw,Chen:2022asf,Meng:2022ozq}.
Many exotic states are near the thresholds of a pair of heavy
hadrons from several to several tens MeVs, which makes the hadronic
molecular picture the popular one.

The hidden-charm pentaquarks were first reported in the
$\Lambda_b^0\rightarrow J/\psi p K^-$ process by the LHCb
Collaboration \cite{Aaij:2015tga,Aaij:2016phn}. In 2019, the updated
analysis with about $10$ times larger statistics showed that there
exist three pronounced states $P_{c}(4312)$, $P_{c}(4440)$, and
$P_c(4457)$ in the $J/\psi p$ invariant mass
spectrum~\cite{Aaij:2019vzc}. Since the masses of the reported
pentaquarks are below the thresholds of the $\Sigma_c\bar{D}^{(*)}$
systems, the molecular explanations of these states have been
discussed in lots of
works~\cite{Wu:2010jy,Yang:2011wz,Wang:2011rga,Chen:2019asm,Liu:2019tjn,He:2019ify,Xiao:2019aya,Meng:2019ilv,Yamaguchi:2019seo,PavonValderrama:2019nbk,Chen:2019bip,Burns:2019iih,Du:2019pij,Wang:2019ato}.
After the observation of $P_c$ states, their strange partners were
studied in
Refs.~\cite{Chen:2016ryt,Santopinto:2016pkp,Shen:2019evi,Xiao:2019gjd,Wang:2019nvm,Chen:2015sxa}.
The $P_{cs}$ states were experimentally investigated in the
$\Xi_b^-\rightarrow J/\psi \Lambda K^-$ process by the LHCb
\cite{LHCb:2020jpq}. They reported the evidence of a new pentaquark
state $P_{cs}(4459)$ with $3\sigma$ significance. Besides, the LHCb
Collaboration \cite{LHCb:2021chn} also showed the evidence of a
$P_c(4337)^+$ signal from their four-dimensional amplitude analysis
in the $B_s^0\rightarrow J/\psi p\bar{p}$ process with less than
$4\sigma$ significance. Very recently, the LHCb Collaboration
reported a very narrow structure $T_{cc}^+(3875)$ in the $D^0D^0\pi^+$
invariant spectrum~\cite{LHCb:2021auc,LHCb:2021vvq}. This state lies
slightly below the $D^{*+}D^0$ threshold by about $300$ keV.
Before the observation of $T_{cc}^+(3875)$, the $QQ\bar{q}\bar{q}$
configurations had been studied by many works based on the tetraquark
picture~\cite{Carlson:1987hh,Gelman:2002wf,Vijande:2003ki,Cui:2006mp,Navarra:2007yw,Vijande:2007rf,Ebert:2007rn,Lee:2009rt,Yang:2009zzp,Wang:2017uld,Du:2012wp,Luo:2017eub,Karliner:2017qjm,Eichten:2017ffp} or the molecular scenario~\cite{Sakai:2017avl,Manohar:1992nd,Pepin:1996id,Janc:2004qn,Ikeda:2013vwa,Carames:2011zz,Molina:2010tx,Li:2012ss,Feng:2013kea,Junnarkar:2018twb,Maiani:2019lpu,Wang:2018atz,Liu:2019stu}. 

It is natural to investigate the existence of molecular states that
are composed of two baryons. Up to now, the deuteron is the only
well-established molecule composed of a proton and a neutron.
Although the $\Lambda\Lambda$ di-baryon state was predicted long ago
by Jaffe~\cite{Jaffe:1976yi} and received amounts of
attentions~\cite{Balachandran:1983dj,Takahashi:2001nm,Polinder:2007mp,Yoon:2007aq,Inoue:2010es,NPLQCD:2010ocs,Morita:2014kza,Li:2018tbt},
the existence of this state is still controversial. The di-baryon
systems that are composed of two charmed baryons are more likely to
be bound, since the large reduced mass can facilitate the
stabilization of such systems. As the heavy flavor siblings of the
nucleon, the molecular states in the $\Sigma_c\Sigma_c$ system have
been investigated in a series of
works~\cite{Lee:2011rka,Garcilazo:2020acl,Gerasyuta:2011zx,Froemel:2004ea,Xia:2021tof,Dong:2021bvy,Ling:2021asz}.
In this work, we will mainly concentrate on the interactions of the
$\Sigma_c\Sigma_c$ system, and explore the relevance of intermediate
states in the two-pion exchange (TPE) loops within the chiral
effective field theory ($\chi$EFT). In addition, we will also study
the interactions of the $\Lambda_c\Lambda_c$ and $\Lambda_c\Sigma_c$
systems. The $\Lambda_c\Lambda_c$ system was studied in various
models~\cite{Meguro:2011nr,Lee:2011rka,Huang:2013rla,Carames:2015sya,Garcilazo:2020acl,Li:2012bt,Wang:2021qmn,Gerasyuta:2011zx,Xia:2021tof,Chen:2017vai,Lu:2017dvm}.
It was shown that the single-channel $\Lambda_c\Lambda_c$ cannot
form the bound
state~\cite{Lee:2011rka,Garcilazo:2020acl,Wang:2021qmn,Xia:2021tof,Dong:2021bvy,Chen:2017vai,Gerasyuta:2011zx},
while in Refs.
\cite{Meguro:2011nr,Huang:2013rla,Li:2012bt,Lu:2017dvm} the authors
argued that the coupling to the attractive
$\Sigma_c^{(*)}\Sigma_c^{(*)}$ channels may lead the
$\Lambda_c\Lambda_c$ system to be bound. The $\Lambda_c\Sigma_c$
molecules were studied in the one-boson-exchange
model~\cite{Li:2012bt} and the dispersion relation
technique~\cite{Gerasyuta:2011zx}, while the $\Lambda_c\Sigma_c$
bound states were disfavored in Ref.~\cite{Xia:2021tof}. The
calculations and discussions on the $\Lambda_c\Sigma_c$ and
$\Lambda_c\Lambda_c$ interactions are relegated to the
appendix~\ref{app2} in this work.

The chiral effective field theory ($\chi$EFT) has been widely
applied to describe the nuclear forces (see
\cite{Bernard:1995dp,Epelbaum:2008ga,Machleidt:2011zz,Meissner:2015wva,Hammer:2019poc}
for reviews), as well as the $D_{s}(2317)$ \cite{Huang:2021fdt},
$T_{cc}$ \cite{Xu:2017tsr}, $P_c$ \cite{Meng:2019ilv,Wang:2019ato},
and $P_{cs}$ \cite{Wang:2019nvm} states (see the recent
review~\cite{Meng:2022ozq}). In the framework of $\chi$EFT, we
include the leading order (LO) contact term, one-pion exchange
(OPE), and two-pion exchange (TPE) contributions to account for the
short-, long-, and intermediate-range interactions of the doubly
charmed di-baryon systems, respectively. Among them, the OPE and TPE
interactions (loops are calculated with the dimensional
regularization scheme) can be derived from the chiral Lagrangians,
and their contributions are definite since the involved coupling
constants can be determined from experiments.  The meson-meson
$(\bar{c}q)$-$(\bar{c}q)$, baryon-meson $(cqq)$-$(\bar{c}q)$, and
baryon-baryon $(cqq)$-$(cqq)$ systems are composed of (anti)charm
quarks and light quarks (without light antiquarks), thus the
exchanged light currents may play very similar dynamic roles in the
heavy quark limit~\cite{Chen:2021cfl,Chen:2021spf}. Therefore, for
the undetermined short-range contact interactions, they can be
related to each other via a quark level Lagrangian
\cite{Meng:2019nzy,Wang:2019nvm,Wang:2020dhf,Chen:2021cfl,Chen:2021spf}
since they are required to obey the SU(3) flavor symmetry and heavy
quark symmetry.

This paper is organized as follows. In Sec.~\ref{sec2}, we present
the chiral effective Lagrangians and the effective potentials. In
Sec.~\ref{sec3}, we present our numerical results and discussions.
In Sec.~\ref{sec4}, we conclude this work with a short summary. We
present the results for the $\Lambda_c\Lambda_c$ and
$\Lambda_c\Sigma_c$ systems in appendix~\ref{app2}.

\section{Chiral effective Lagrangians and effective potentials}\label{sec2}

\subsection{Chiral effective Lagrangians}
In the SU(2) case, the light diquark in the ground-state singly
charmed baryons can be the antisymmetric isosinglet or symmetric
isotriplet. The corresponding total spin of the light diquark
component is $j_{l}=0$ or $j_{l}=1$. The spin-$\frac{1}{2}$
isosinglet is defined as
\begin{eqnarray}
\psi_1=\left[
  \begin{array}{cc}
    0 & \Lambda_c^+ \\
    -\Lambda_c^+ & 0 \\
  \end{array}
\right],\quad
\end{eqnarray}
while the spin-$\frac{1}{2}$ and spin-$\frac{3}{2}$ isospin triplets
are defined as
\begin{eqnarray}
\psi_3=\left[
  \begin{array}{cc}
    \Sigma_c^{++} & \frac{\Sigma_c^+}{\sqrt{2}} \\
    \frac{\Sigma_c^+}{\sqrt{2}} & \Sigma_c^0\\
  \end{array}
\right],\quad \psi^{\mu}_{3^*}=\left[
  \begin{array}{cc}
    \Sigma_c^{*++} & \frac{\Sigma_c^{*+}}{\sqrt{2}} \\
    \frac{\Sigma_c^{*+}}{\sqrt{2}} & \Sigma_c^{*0} \\
  \end{array}
\right]^{\mu}.
\end{eqnarray}
In the heavy baryon reduction formalism~\cite{Scherer:2002tk}, the
heavy baryon field can be decomposed into the light and heavy
components through the projection operators $(1\pm\slashed{v})/2$,
\begin{eqnarray}
\label{B} \mathcal{B}_i=e^{iM_iv\cdot
x}\frac{1+\slashed{v}}{2}\psi_i,\quad \label{H}
\mathcal{H}_i=e^{iM_iv\cdot x}\frac{1-\slashed{v}}{2}\psi_i,
\end{eqnarray}
with $\psi_i$ the heavy baryon fields $\psi_1$, $\psi_3$,
$\psi_{3^*}^\mu$, and $M_i$ the masses of heavy baryons. The
four-velocity $v_\mu=(1,\bm{0})$ in the rest frame of the heavy
baryons. In the leading order expansion, only the light component
$\mathcal{B}_i$ is kept.

Then we introduce the leading order chiral Lagrangians to describe
the interactions between the charmed baryons and pion
fields~\cite{Cheng:1992xi,Liu:2012uw}
\begin{widetext}
\begin{eqnarray}
\mathcal{L}_{\cal
B\varphi}&=&\frac{1}{2}{\rm{Tr}}\left[\mathcal{B}_1\left(i v\cdot
D\right)\mathcal{B}_1\right]+2g_2{\rm{Tr}}\left(\bar{\mathcal{B}}_3\mathcal{S}\cdot
u\mathcal{B}_1+\text{H.c.}\right)
+g_3\text{Tr}\left(\bar{B}_{3^*}^\mu u_\mu\mathcal{B}_3+\text{H.c.}\right)+g_4\text{Tr}\left(\bar{\mathcal{B}}_{3^*}^\mu u_{\mu}\mathcal{B}_1+\text{H.c.}\right)\nonumber\\
&&+\text{Tr}\left[\bar{\mathcal{B}}_3\left(iv\cdot
D-\delta_c\right)\mathcal{B}_3\right]+2g_5\text{Tr}\left(\bar{\mathcal{B}}_{3^*}^\mu\mathcal{S}\cdot
u\mathcal{B}_{3^*}^\mu\right)
-\text{Tr}\left[\bar{\mathcal{B}}_{3^*}^\mu\left(iv\cdot
D-\delta_d\right)\mathcal{B}_{3*\mu}\right]+2g_1\text{Tr}\left(\bar{\mathcal{B}}_3\mathcal{S}\cdot
u\mathcal{B}_3\right), \label{BpLag}
\end{eqnarray}
\end{widetext}
where $\mathcal{S}^{\mu}=\frac{i}{2}\gamma_5\sigma^{\mu\nu}v_\nu$ is
the covariant spin operator for the spin-$\frac{1}{2}$ baryon. The
covariant derivative
$D_\mu\psi=\partial_\mu\psi+\Gamma_\mu\psi+\psi\Gamma_\mu^T$, with
$\Gamma_\mu^T$ the transpose of $\Gamma_\mu$. The chiral connection
$\Gamma_\mu$ and axial-vector current $u_\mu$ are defined as
\begin{eqnarray}\label{GU}
\Gamma_\mu=\frac{1}{2}\left[\xi^\dagger,\partial_\mu
\xi\right],\qquad
u_\mu=\frac{i}{2}\left\{\xi^\dagger,\partial_\mu\xi\right\},
\end{eqnarray}
with
\begin{eqnarray}
\xi^2=U=\text{exp}\left(\frac{i\varphi}{f_\pi}\right), \quad
\varphi=\left[
 \begin{array}{cc}
\pi^0 & \sqrt{2}\pi+ \\
\sqrt{2}\pi^- & -\pi^0 \\
 \end{array}
 \right],
\end{eqnarray}
in which $f_\pi=92.4$ MeV is the pion decay constant.

The coupling constants $g_2=-0.6$ and $g_4=1.04$ in Eq.
(\ref{BpLag}) are determined from the partial decay widths of the
$\Sigma_c\rightarrow \Lambda_c\pi$ and $\Sigma_c^*\rightarrow
\Lambda_c\pi$~\cite{ParticleDataGroup:2020ssz}, respectively. The
other coupling constants $g_1=0.98$, $g_3=0.85$, and $g_5=-1.47$ are
obtained by relating them to $g_2$ via the quark
model~\cite{Meguro:2011nr,Liu:2011xc,Meng:2018gan}. The mass
splittings $\delta_a=M_{\Sigma^*_c}-M_{\Sigma_c}=65$ MeV,
$\delta_b=M_{\Sigma_c}-M_{\Lambda_c}=168.5$ MeV, and
$\delta_c=M_{\Sigma_c^*}-M_{\Lambda_c}=233.5$ MeV are extracted from
the masses of the $\Lambda_c$, $\Sigma_c$, and $\Sigma_c^*$
baryons~\cite{ParticleDataGroup:2020ssz}.

The following Lagrangian is constructed to describe the four-body
contact interactions (FBCI) among $\mathcal{B}_1$, $\mathcal{B}_3$
and $\mathcal{B}_3^\ast$,
\begin{widetext}
\begin{eqnarray}
\mathcal{L}_{\cal B\cal
B}&=&C_a\text{Tr}\left(\bar{\mathcal{B}}_1\mathcal{B}_1\right)\text{Tr}\left(\bar{\mathcal{B}}_1\mathcal{B}_1\right)+C_b\text{Tr}\left(\bar{\mathcal{B}}_1\mathcal{B}_1\right)\text{Tr}\left(\bar{\psi}^\mu\psi_\mu\right)
+C_c\text{Tr}\left(\bar{\psi}^\mu\psi_\mu\right)\text{Tr}\left(\bar{\psi}^\nu\psi_\nu\right)
+D_a\text{Tr}\left(\bar{\psi}^\mu\tau_i\psi_\mu\right)\text{Tr}\left(\bar{\psi}^\nu\tau^{i}\psi_\nu\right)\nonumber\\
&&+iC_d\epsilon_{\mu\nu\alpha\beta}v^\alpha\text{Tr}\left(\bar{\psi}^\rho\gamma^\beta\gamma_5\psi_\rho\right)\text{Tr}\left(\bar{\psi}^\mu\psi^\nu\right)
+iD_b\epsilon_{\mu\nu\alpha\beta}v^{\alpha}\text{Tr}\left(\bar{\psi}^\rho\gamma^\beta\gamma_5\tau^i\psi_\rho\right)\text{Tr}\left(\bar{\psi}^\mu\tau_i\psi^\nu\right),\label{FBCI}
\end{eqnarray}
where we use the super-field notation for the heavy quark spin
doublet $(\mathcal{B}_3,\mathcal{B}_{3^\ast})$,
\begin{eqnarray}
\psi^{\mu}=\mathcal{B}_{3^*}^\mu-\frac{1}{\sqrt{3}}\left(\gamma^\mu+v^\mu\right)\gamma^5\mathcal{B}_3,~\text{and
}\bar{\psi}^\mu=\bar{\mathcal{B}}_{3^*}^\mu+\frac{1}{3}\bar{\mathcal{B}}_3\gamma^5\left(\gamma^\mu+v^{\mu}\right).
\end{eqnarray}
The $C_a$, $C_b$, $C_c$, $D_a$, $C_d$, and $D_b$ are six independent
low-energy constants (LECs) that will be determined later. $\tau^i$
represents the isospin Pauli matrix. Note that the other forms of
the LO contact Lagrangians with different Lorentz structures are
also allowed, but they can be expressed as the linear combinations
of the terms involved in Eq. (\ref{FBCI}).
\end{widetext}

\subsection{Effective potentials}

Due to the symmetry constraint for the identical
particles~\cite{Chen:2021cfl}, the two $\Sigma_c$ baryons can form
\begin{eqnarray}
[\Sigma_c\Sigma_c]_{J}^I, \text{ with }(J,I)=(0,0),~(1,1),~(0,2).
\end{eqnarray}
In what follows the subscript and superscript denote the total spin and isospin of the corresponding di-baryon systems, respectively. 

The scattering amplitude $\mathcal{M}(\bm{q})$ can be calculated by
expanding the Lagrangians in Eqs.~\eqref{BpLag}-\eqref{FBCI}. It can
be related to the effective potential under the Breit approximation,
i.e.,
\begin{eqnarray}
\mathcal{V}\left(\bm{q}\right)=-\frac{\mathcal{M}(\bm{q})}{\sqrt{2M_12M_22M_32M_4}},\label{Breit}
\end{eqnarray}
where the $M_{1,2}$ and $M_{3,4}$ are the masses of the incoming and outgoing particles, respectively. 

The expressions for the LO contact and OPE potentials read
\begin{eqnarray}
\mathcal{V}^{\mathrm{ct}}_{\Sigma_c\Sigma_c}&=&-4\left[C_c+\left(\bm{I}_1\cdot\bm{I}_2\right)D_a\right]\nonumber\\&&
+\frac{8}{9}\left[C_d+\left(\bm{I}_1\cdot\bm{I}_2\right)D_b\right]\bm{\sigma}_1\cdot\bm{\sigma}_2,\label{SSCon}\\
\mathcal{V}^{\mathrm{OPE}}_{\Sigma_c\Sigma_c}&=&-\left(\bm{I}_1\cdot\bm{I}_2\right)\frac{g_1^2}{4f_\pi^2}\frac{\left(\bm{\sigma}_1\cdot\bm{q}\right)\left(\bm{\sigma}_2\cdot\bm{q}\right)}{\bm{q}^2+m_\pi^2},\label{SSOpe}
\end{eqnarray}
in which the eigenvalues of the $\bm{I}_1\cdot\bm{I}_2$ and
$\bm{\sigma}_1\cdot\bm{\sigma}_2$ operators can be calculated with
the following equations
\begin{eqnarray}
\left\langle\bm{I}_1\cdot\bm{I}_2 \right\rangle&=&\frac{1}{2}\left[I\left(I+1\right)-I_1\left(I_1+1\right)-I_2\left(I_2+1\right)\right],\nonumber\\
\left\langle\bm{\sigma}_1\cdot\bm{\sigma}_2\right\rangle&=&2\left[S\left(S+1\right)-S_1\left(S_1+1\right)-S_2
\left(S_2+1\right)\right].\nonumber
\end{eqnarray}
As shown in Eq.~(\ref{SSCon}), the contact potential of the
$\Sigma_c\Sigma_c$ system consists of four parts---the central term,
the isospin-isospin interaction term $\bm{I}_1\cdot\bm{I}_2$, the
spin-spin interaction term $\bm{\sigma}_1\cdot\bm{\sigma}_2$, and
the isospin-spin interaction coupled term
$(\bm{I}_1\cdot\bm{I}_2)(\bm{\sigma}_1\cdot\bm{\sigma}_2)$. The
contact and OPE potentials for the different $\Sigma_c\Sigma_c$
states can be distinguished with the matrix elements of the isospin
and spin operators. In the $S$-wave case, the operator
$(\bm{\sigma}_1\cdot\bm{q})(\bm{\sigma}_2\cdot\bm{q})$ can be
simplified with the following
replacement~\cite{Sun:2012zzd,Li:2012ss}
\begin{eqnarray}
\left(\bm{\sigma}_1\cdot\bm{q}\right)\left(\bm{\sigma}_2\cdot\bm{q}\right)\rightarrow
\frac{1}{3}\bm{q}^2\left(\bm{\sigma}_1\cdot\bm{\sigma}_2\right).\label{qqapprox}
\end{eqnarray}

\begin{figure*}[htbp]
\includegraphics[width=0.7\linewidth]{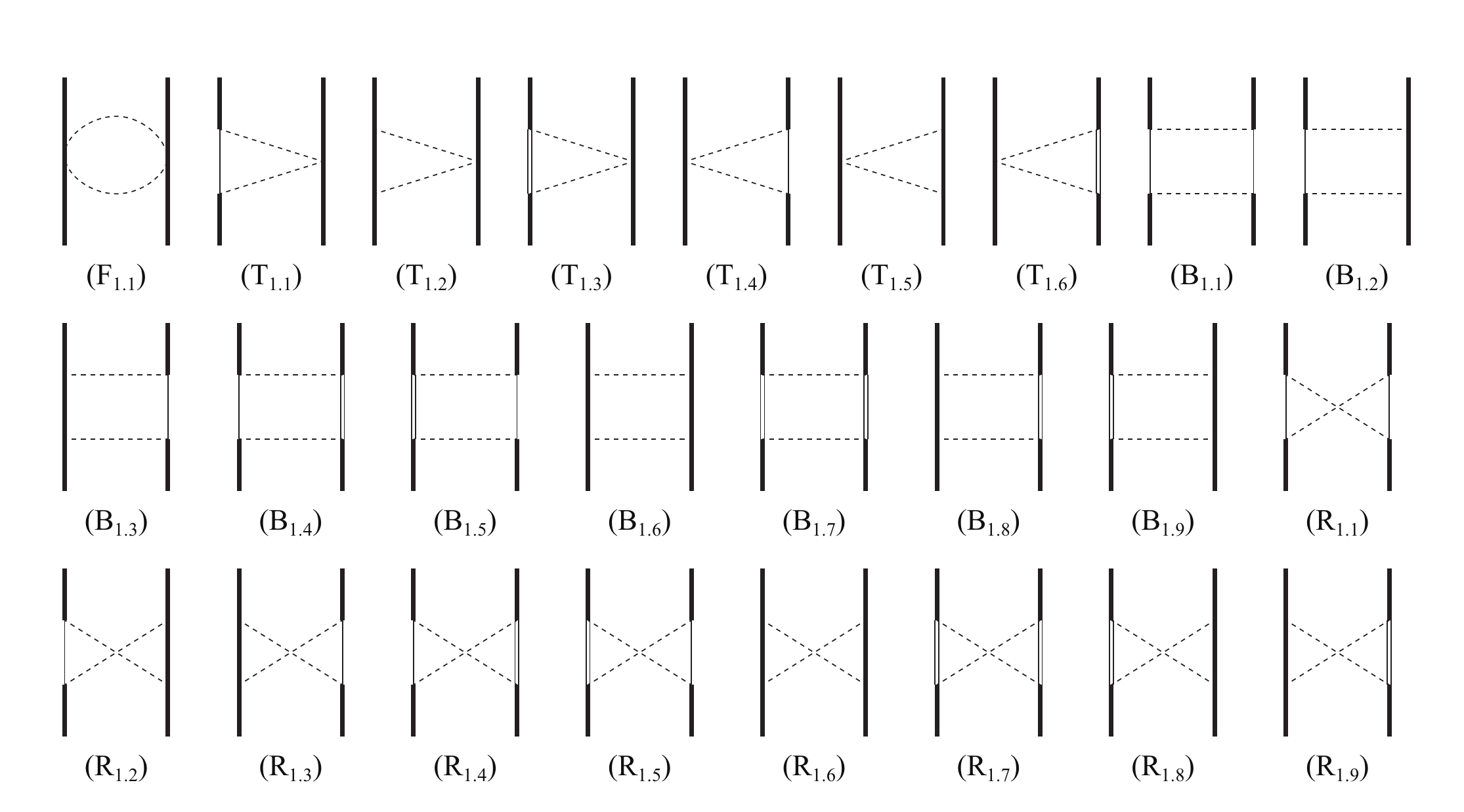}
\caption{TPE diagrams for the $\Sigma_c\Sigma_c$ system at NLO. We
use the thin line, thick line, double-thin line, and dashed line to
denote the $\Lambda_c$, $\Sigma_c$, $\Sigma_c^*$, and $\pi$,
respectively.}\label{sigcsigctwopionfig}
\end{figure*}

The NLO TPE diagrams for the $\Sigma_c\Sigma_c$ system are
illustrated in Fig.~\ref{sigcsigctwopionfig}. The TPE diagrams
include the football diagram ($\mathrm{F}_{1.1}$), triangle diagrams
($\mathrm{T}_{1.1}$)-($\mathrm{T}_{1.6}$), box diagrams
($\mathrm{B}_{1.1})$-$(\mathrm{B}_{1.9})$, and cross diagrams
$(\mathrm{R}_{1.1})$-$(\mathrm{R}_{1.9})$.
The analytical expressions of the TPE potentials from these diagrams
can be collectively written as
\begin{widetext}
\begin{eqnarray}
\mathcal{V}^{\mathrm{F}_{1.1}}&=&\left(\bm{I}_1\cdot\bm{I}_2\right)\frac{1}{f_\pi^4}J_{22}^F,\label{F11}\\
\mathcal{V}^{\mathrm{T}_{1.j}}&=&\left(\bm{I}_1\cdot\bm{I}_2\right)\frac{\mathcal{C}^{\mathrm{T}_{1.j}}}{f_\pi^4}\left[\bm{q}^2C_1^{\mathrm{T}_{1.j}}\left(J_{24}^T+J_{33}^T\right)+C_2^{\mathrm{T}_{1.j}}J_{34}^T\right](\mathcal{E}^{\mathrm{T}_{1.j}}),\label{T1j}\\
\mathcal{V}^{\mathrm{B}_{i.j}}&=&F^{\mathrm{B}_{i.j}}_I\frac{\mathcal{C}^{\mathrm{B}_{i.j}}}{f_\pi^4}\Big[\bm{q}^2C_1^{\mathrm{B}_{i.j}}J_{21}^B+\frac{2}{3}\bm{q}^2C_2^{\mathrm{B}_{i.j}}\left(\bm{\sigma}_1\cdot\bm{\sigma}_2\right)J_{21}^B+\bm{q}^4C_3^{\mathrm{B}_{i.j}}\left(J_{22}^B+2J_{32}^B+J_{43}^B\right)+C_5^{\mathrm{B}_{i.j}}J_{41}^B \nonumber\\ &&+\bm{q}^2C_4^{\mathrm{B}_{i.j}}\left(J_{31}^B+J_{42}^B\right)\Big](\mathcal{E}_1^{\mathrm{B}_{i.j}},\mathcal{E}_2^{\mathrm{B}_{i.j}}),\label{Bij}\\
\mathcal{V}^{\mathrm{R}_{i.j}}&=&F^{\mathrm{R}_{i.j}}_I\frac{\mathcal{C}^{\mathrm{R}_{i.j}}}{f_\pi^4}\left[\bm{q}^2C_1^{\mathrm{R}_{i.j}}J_{21}^R+\frac{2}{3}\bm{q}^2C_2^{\mathrm{R}_{i.j}}\left(\bm{\sigma}_1\cdot\bm{\sigma}_2\right)J_{21}^R
+\bm{q}^4C_3^{\mathrm{R}_{i.j}}\left(J_{22}^R+2J_{32}^R+J_{43}^R\right)+C_5^{\mathrm{R}_{i.j}}J_{41}^R\right.\nonumber\\&&\left.+\bm{q}^2C_4^{\mathrm{R}_{i.j}}\left(J_{31}^R+J_{42}^R\right)\right]
(\mathcal{E}_1^{\mathrm{R}_{i.j}},\mathcal{E}_2^{\mathrm{R}_{i.j}}),\label{Rij}
\end{eqnarray}
\end{widetext}
in which we have used Eq.~(\ref{qqapprox}) in
Eqs.~\eqref{Bij}-\eqref{Rij}. The coefficients defined in
Eq.~(\ref{T1j}) for the triangle diagrams are collected in Table
\ref{T1j factor}. The $\Sigma_c\Sigma_c$ system can couple to
isospin $I=0,1,2$. The isospin factors $F_{I}^{\mathrm{B}_{i.j}}$
($F_{I}^{\mathrm{R}_{i.j}}$) of the box (cross) diagrams defined in
Eq.~(\ref{Bij}) [Eq. (\ref{Rij})] are collected in Table~\ref{iso
factor}. The other coefficients defined in Eq.~(\ref{Bij}) and
Eq.~(\ref{Rij}) are given in Table~\ref{BR factor}.

The scalar loop functions $J_{ab}^F$, $J_{ab}^T$, $J_{ab}^B$, and $J_{ab}^R$ defined in Eqs.~\eqref{F11}-\eqref{Rij} have been given in the Appendix of Refs.~\cite{Meng:2019ilv,Wang:2019ato}. The residual energies $\mathcal{E}^{\mathrm{T}_{i.j}}$, $\mathcal{E}_{1(2)}^{\mathrm{B}_{i.j}}$, $\mathcal{E}_{1(2)}^{\mathrm{R}_{i.j}}$ are defined as the mass differences of the incoming and intermediate states.

\begin{table}
\renewcommand\arraystretch{1.5}
\caption{The coefficients defined in Eq.~(\ref{T1j}) for the
$\Sigma_c\Sigma_c$ system. \label{T1j factor}}
 \setlength{\tabcolsep}{4.6mm}
{
\begin{tabular}{c|cccccccccccccc}
\toprule[0.7pt]
&$\mathcal{C}^{\mathrm{T}_{1.j}}$&$C_{1}^{\mathrm{T}_{1.j}}$&$C_2^{\mathrm{T}_{1.j}}$&$\mathcal{E}^{\mathrm{T}_{1.j}}$\\
\hline
$j=1,4$&$\frac{g_2^2}{2}$&1&3&$\delta_b$\\
$j=2,5$&$\frac{g_1^2}{4}$&1&3&0\\
$j=3,6$&$\frac{g_3^2}{6}$&1&3&$-\delta_a$\\
\bottomrule[0.7pt]
\end{tabular}}
\end{table}

\begin{table}
\centering
\renewcommand\arraystretch{1.5}
\caption{Isospin factors of the box and cross diagrams for the
$\Sigma_c\Sigma_c$ system defined in Eq. (\ref{Bij}) and Eq.
(\ref{Rij}). \label{iso factor}} \setlength{\tabcolsep}{1.5mm} {
\begin{tabular}{c|cccccccccccccc}
\toprule[0.7pt]
&$j=1$&$j=2,3,4,5$&$j=6,7,8,9$\\
\hline
$[F_{2}^{\mathrm{B}_{1.j}},F_{1}^{\mathrm{B}_{1.j}},F_{0}^{\mathrm{B}_{1.j}}]$&$[0,0,3]$&$[0,2,0]$&$[1,1,4]$\\
$[F_{2}^{\mathrm{R}_{1.j}},F_{1}^{\mathrm{R}_{1.j}},F_{0}^{\mathrm{R}_{1.j}}]$&$[1,-1,1]$&$[1,1,-2]$&$[2,0,2]$\\
\bottomrule[0.7pt]
\end{tabular}}
\end{table}

\begin{table*}
\centering
\renewcommand\arraystretch{1.5}
\caption{The coefficients defined in Eq.~(\ref{Bij}) and
Eq.~(\ref{Rij}) for the $\Sigma_c\Sigma_c$ system. \label{BR
factor}} \setlength{\tabcolsep}{3.0mm} {
\begin{tabular}{cl|ccccccccccccc}
\toprule[0.7pt]
&&$\mathcal{C}^{(\mathrm{B}/\mathrm{R})_{i.j}}$&$C_1^{(\mathrm{B}/\mathrm{R})_{i.j}}$&$C_2^{(\mathrm{B}/\mathrm{R})_{i.j}}$&$C_3^{(\mathrm{B}/\mathrm{R})_{i.j}}$
&$C_4^{(\mathrm{B}/\mathrm{R})_{i.j}}$&$C_5^{(\mathrm{B}/\mathrm{R})_{i.j}}$&$\mathcal{E}_1^{(\mathrm{B}/\mathrm{R})_{i.j}}$&$\mathcal{E}_2^{(\mathrm{B}/\mathrm{R})_{i.j}}$\\
\hline
\multirow{6}*{$i=1$}&$j=1$&$\frac{g_2^4}{4}$&$-1$&$1/-1$&$1$&$-10$&$15$&$\delta_b$&$\delta_b$\\
&$j=2,3$&$\frac{g_1^2g_2^2}{8}$&$-1$&$1/-1$&$1$&$-10$&$15$&$\delta_b$&0\\
&$j=4,5$&$\frac{g_2^2g_3^2}{24}$&$-2$&$-1/1$&$2$&$-20$&$30$&$\delta_b$&$-\delta_a$\\
&$j=6$&$\frac{g_1^4}{16}$&$-1$&$1/-1$&$1$&$-10$&$15$&0&0\\
&$j=7$&$\frac{g_3^4}{144}$&$-4$&$1/-1$&$4$&$-40$&$60$&$-\delta_a$&$-\delta_a$\\
&$j=8,9$&$\frac{g_1^2g_3^2}{48}$&$-2$&$-1/1$&2&$-20$&30&0&$-\delta_a$\\
\bottomrule[0.7pt]
\end{tabular}}
\end{table*}

\section{Numerical results and discussions}\label{sec3}
\subsection{Contact terms}

Since the experimental data or lattice QCD simulations for the
interactions of the doubly charmed di-baryon systems are still
absent, we have to use a practical way to estimate the LECs in
Eq.~\eqref{FBCI}. In
Refs.~\cite{Meng:2019nzy,Wang:2019nvm,Wang:2020dhf,Chen:2021cfl,Chen:2021spf},
we proposed to bridge the unknown LECs to the systems with
experimental data via a quark-level Lagrangian. This effective
approach has been successfully used to predict the $P_{cs}$
states~\cite{Wang:2019nvm}. In this approach, the contact
interactions of the heavy flavor di-hadron systems are ascribed to
the exchange of the light meson currents. The interactions induced
by the light meson exchanges dominate the interactions of the heavy
flavor di-hadron systems. The light quark components of the
experimentally observed $P_c$ states and the $\Sigma_c\Sigma_c$ are
all light quarks (without light antiquarks). Thus, the exchanged
light mesons shall play very similar dynamic roles in the
hidden-charm meson-baryon and double-charm di-baryon
systems~\cite{Chen:2021cfl,Chen:2021spf}. Therefore, we can use the
data of the $P_c$ states as input to estimate the LECs of the
$\Sigma_c\Sigma_c$ system.

In Ref.~\cite{Wang:2019ato}, we presented a detailed study on the
interactions of the $\Sigma_c^{(*)}\bar{D}^{(*)}$ systems with
$\chi$EFT. The LO contact terms of the
$[\Sigma_c\bar{D}^*]_{1/2}^{1/2}$ and
$[\Sigma_c\bar{D}^*]^{1/2}_{3/2}$ systems are given
as~\cite{Wang:2019ato}
\begin{eqnarray}
\mathcal{V}_{[\Sigma_c\bar{D}^*]_{1/2}^{1/2}}&=&-\mathbb{D}_1-\frac{4}{3}\mathbb{D}_2,\label{HSD1212}\\
\mathcal{V}_{[\Sigma_c\bar{D}^*]_{3/2}^{1/2}}&=&-\mathbb{D}_1+\frac{2}{3}\mathbb{D}_2,\label{HSD3212}
\end{eqnarray}
where we adopt the same notations as that of
Ref.~\cite{Wang:2019ato}. The $\mathbb{D}_1$ and $\mathbb{D}_2$ are
two LECs denoting the strength of the central potential and
spin-spin interaction, respectively. Two sets of solutions for
$\mathbb{D}_1$ and $\mathbb{D}_2$ were obtained in
Ref.~\cite{Wang:2019ato} via fitting the binding energies of $P_c$
states, here, we adopt
\begin{eqnarray}
\Lambda&=&0.5~\text{GeV},\quad
\begin{cases}\label{eq:LECs5}
    \mathbb{D}_1=52.0~\textrm{GeV}^{-2},\\
     \mathbb{D}_2=-4.0~\textrm{GeV}^{-2},
\end{cases}
\end{eqnarray}
where $\Lambda$ is the cutoff that will be introduced in
Eq.~\eqref{gaussform}. On the other hand, the expressions of the
quark-level contact terms for the $[\Sigma_c\bar{D}^*]_{1/2}^{1/2}$
and $[\Sigma_c\bar{D}^*]^{1/2}_{3/2}$ can be written as
\begin{eqnarray}
\mathcal{V}_{[\Sigma_c\bar{D}^*]_{1/2}^{1/2}}&=&-\frac{10}{3}\tilde{g}_s+\frac{40}{9}\tilde{g}_a,\label{QSD1212}\\
\mathcal{V}_{[\Sigma_c\bar{D}^*]_{3/2}^{1/2}}&=&-\frac{10}{3}\tilde{g}_s-\frac{20}{9}\tilde{g}_a,\label{QSD3212}
\end{eqnarray}
in which the notations for the quark-level couplings $\tilde{g}_s$
and $\tilde{g}_a$ are the same as those of Ref. \cite{Chen:2021cfl}.
One can easily obtain the values of $\tilde{g}_s$ and $\tilde{g}_a$
through matching Eqs.~\eqref{QSD1212},~\eqref{QSD3212} and
Eqs.~\eqref{HSD1212},~\eqref{HSD3212}.
In Ref.~\cite{Chen:2021cfl}, we only used the LO contact terms to
model the effective potentials of the $\Sigma_c^{(*)}\bar{D}^{(*)}$
systems. In this case, the estimated $\tilde{g}_s$ and $\tilde{g}_a$
are different from those in Ref.~\cite{Wang:2019nvm}, which implies
the LECs receive considerable corrections after we include the
explicit chiral dynamics, e.g., the OPE and TPE interactions.

Then we can use the determined $\tilde{g}_s$ and $\tilde{g}_a$ to
estimate the contact potentials of the $\Sigma_c\Sigma_c$ systems.
Their quark-level $S$-wave contact interactions have been
systematically studied in Ref.~\cite{Chen:2021cfl}, and we have

\begin{eqnarray}
\mathcal{V}_{[\Sigma_c\Sigma_c]_0^0}&=&-\frac{20}{3}\tilde{g}_s+\frac{80}{9}\tilde{g}_a,\label{eq:qlss1}\\
\mathcal{V}_{[\Sigma_c\Sigma_c]_0^2}&=&\frac{16}{3}\tilde{g}_s-\frac{64}{9}\tilde{g}_a,\label{eq:qlss2}\\
\mathcal{V}_{[\Sigma_c\Sigma_c]_1^1}&=&-\frac{8}{3}\tilde{g}_s-\frac{32}{27}\tilde{g}_a.\label{eq:qlss3}
\end{eqnarray}
One can obtain the LECs $C_c$, $C_d$, $D_a$ and $D_b$ through
matching Eqs.~\eqref{SSCon} and~\eqref{eq:qlss1}-\eqref{eq:qlss3}.

In order to search for the possible bound states in the
$\Sigma_c\Sigma_c$ system via solving the Schr\"odinger equation, we
perform the Fourier transformation on $\mathcal{V}(\bm{q})$ to get
the effective potential $\mathcal{V}(r)$ in the coordinate space,
\begin{eqnarray}\label{gaussform}
\mathcal{V}(r)=\int\frac{d^3\bm{q}}{(2\pi)^3}e^{-i\bm{q}\cdot\bm{r}}\mathcal{V}(\bm{q})\mathcal{F}(\bm{q}),
\end{eqnarray}
in which the Gaussian form factor $\mathcal{F}({\bm q})={\rm
exp}(-{\bm q}^{2n}/\Lambda^{2n})$ (with $n=2$) is adopted to
regularize the divergence in this
integral~\cite{Ordonez:1995rz,Epelbaum:1999dj}. The cutoff $\Lambda$
is introduced to exclude the hard momentum contributions.

\subsection{Results and discussion}
In this subsection, we present the effective potentials for
$[\Sigma_c\Sigma_c]_0^{0,2}$ and $[\Sigma_c\Sigma_c]_1^1$ systems
with the cutoff and LECs in Eq.~\eqref{eq:LECs5}. We will discuss
the effective potentials of the $\Sigma_c\Sigma_c$ system in three
cases.
\begin{itemize}
  \item
  Case-I: We only consider the $\Sigma_c$ as the intermediate state in the TPE diagrams.
  \item
  Case-II: We consider both the $\Sigma_c$ and $\Sigma_c^*$ as the intermediate states in the TPE diagrams.
  \item
  Case-III: We consider the $\Lambda_c$, $\Sigma_c$, and $\Sigma_c^*$ as the intermediate states in the TPE diagrams.
\end{itemize}

\begin{figure}[htbp]
\includegraphics[width=0.9\linewidth]{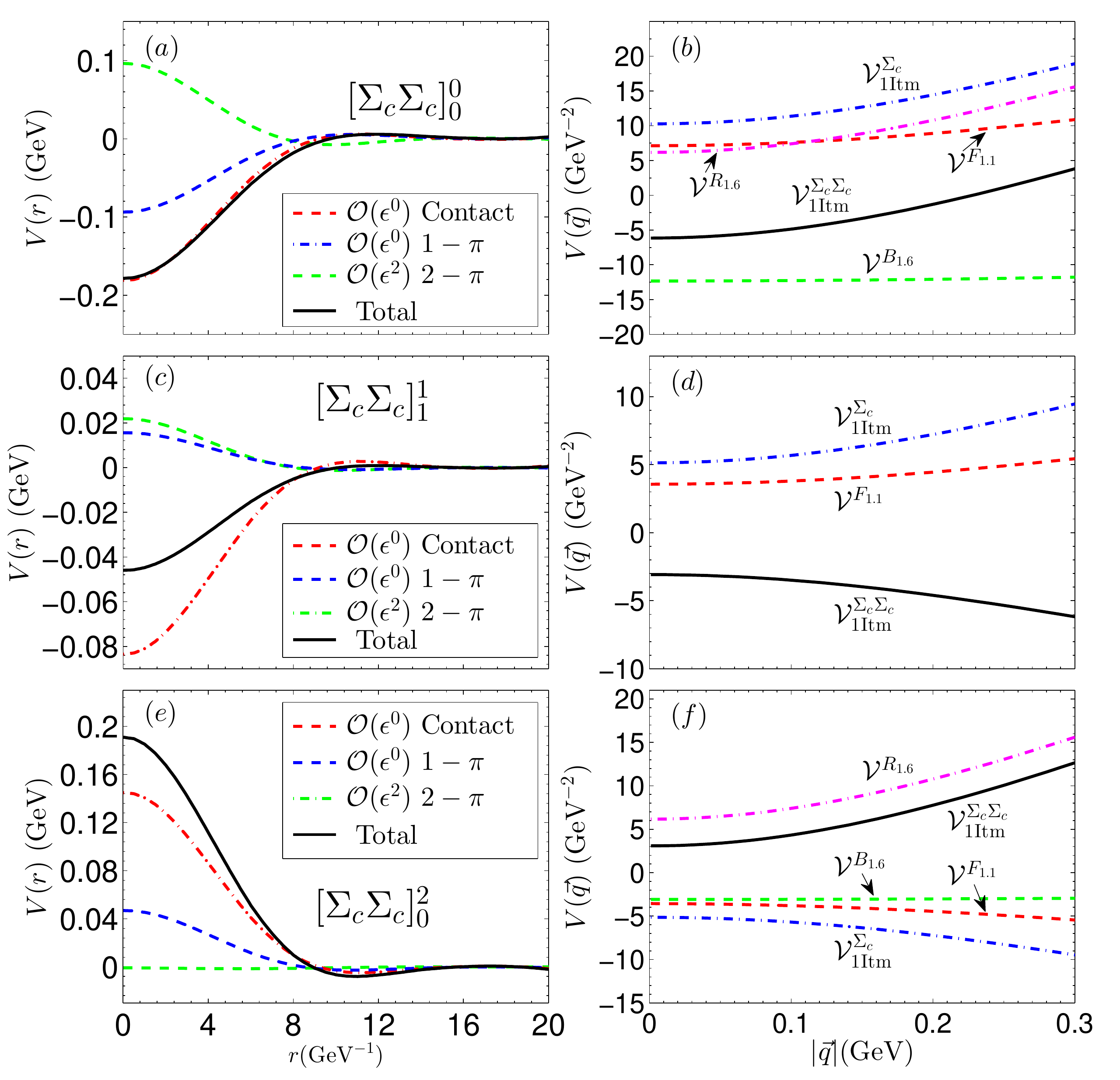}\\
\caption{The first, second, and third rows are the effective
potentials of the $[\Sigma_c\Sigma_c]_0^0$,
$[\Sigma_c\Sigma_c]_1^1$, and $[\Sigma_c\Sigma_c]_0^2$ systems,
respectively. The $\mathcal{V}_{\rm{1Itm}}^{\Sigma_c}$ and
$\mathcal{V}_{\rm{1Itm}}^{\Sigma_c\Sigma_c}$ are defined as
$\mathcal{V}_{\rm{1Itm}}^{\Sigma_c}=\mathcal{V}^{\rm{T}_{1.2}}+\mathcal{V}^{\rm{T}_{1.5}}$
and
$\mathcal{V}_{\rm{1Itm}}^{\Sigma_c\Sigma_c}=\mathcal{V}^{\rm{B}_{1.6}}+\mathcal{V}^{\rm{R}_{1.6}}$,
respectively. \label{sigcsigccase1}}
\end{figure}

In Case-I, only the diagrams $(\mathrm{F}_{1.1})$,
$(\mathrm{T}_{1.2})$, $(\mathrm{T}_{1.5})$, $(\mathrm{B}_{1.6})$,
$(\mathrm{R}_{1.6})$ contribute. The effective potentials from the
contact, OPE, and TPE interactions in coordinate space for the
$[\Sigma_c\Sigma_c]_0^0$, $[\Sigma_c\Sigma_c]_1^1$, and
$[\Sigma_c\Sigma_c]_0^2$ systems are presented in
Figs.~\ref{sigcsigccase1}(a),~\ref{sigcsigccase1}(c),
and~\ref{sigcsigccase1}(e), respectively. From
Fig.~\ref{sigcsigccase1}(a), one sees that for the
$[\Sigma_c\Sigma_c]_0^0$ system, the TPE potential is comparable to
that of OPE but with opposite sign. Consequently, the total
effective potential mainly comes from the contact term, which
provides a large attractive force. The contact potential of the
$[\Sigma_c\Sigma_c]_1^1$~[see Fig.~\ref{sigcsigccase1}(b)] is much
smaller than that of the $[\Sigma_c\Sigma_c]_0^0$ system. Moreover,
the OPE and TPE potentials are all repulsive. Thus, the attractive
force of the $[\Sigma_c\Sigma_c]_1^1$ system is much smaller than
that of the $[\Sigma_c\Sigma_c]_0^0$ system.

In Figs.~\ref{sigcsigccase1}(b),~\ref{sigcsigccase1}(d),
and~\ref{sigcsigccase1}(f), we present the contributions of each
type of the TPE diagrams in momentum space. The
$\mathcal{V}_{\rm{1Itm}}^{\Sigma_c}$ and
$\mathcal{V}_{\rm{1Itm}}^{\Sigma_c\Sigma_c}$ denote
$\mathcal{V}_{\rm{1Itm}}^{\Sigma_c}=\mathcal{V}^{\rm{T}_{1.2}}+\mathcal{V}^{\rm{T}_{1.5}}$,
and
$\mathcal{V}_{\rm{1Itm}}^{\Sigma_c\Sigma_c}=\mathcal{V}^{\rm{B}_{1.6}}+\mathcal{V}^{\rm{R}_{1.6}}$.
The positive (negative) effective potential in momentum space
corresponds to a repulsive (attractive) force in coordinate space.
From Figs.~\ref{sigcsigccase1}(b) and~\ref{sigcsigccase1}(f), one
sees that the contributions of $(\rm{B}_{1.6})$ and $(\rm{R}_{1.6})$
are considerable but with opposite signs. Thus, the sum of these two
potentials give a relatively small
$\mathcal{V}_{1\rm{Itm}}^{\Sigma_c\Sigma_c}$. As collected in
Table~\ref{iso factor}, the isospin factor $F_1^{\rm{R}_{1.6}}$
vanishes for the $[\Sigma_c\Sigma_c]_1^1$ system. Thus, we have
$V_{1\rm{Itm}}^{\Sigma_c\Sigma_c}= V^{\rm{B}_{1.6}}$,
as presented in Fig.~\ref{sigcsigccase1}(d). Besides, the magnitudes of the TPE diagrams are comparable to each other for the $[\Sigma_c\Sigma_c]_0^{0,2}$ and $[\Sigma_c\Sigma_c]_1^1$ systems. 

\begin{figure*}[htbp]
\includegraphics[width=0.6\linewidth]{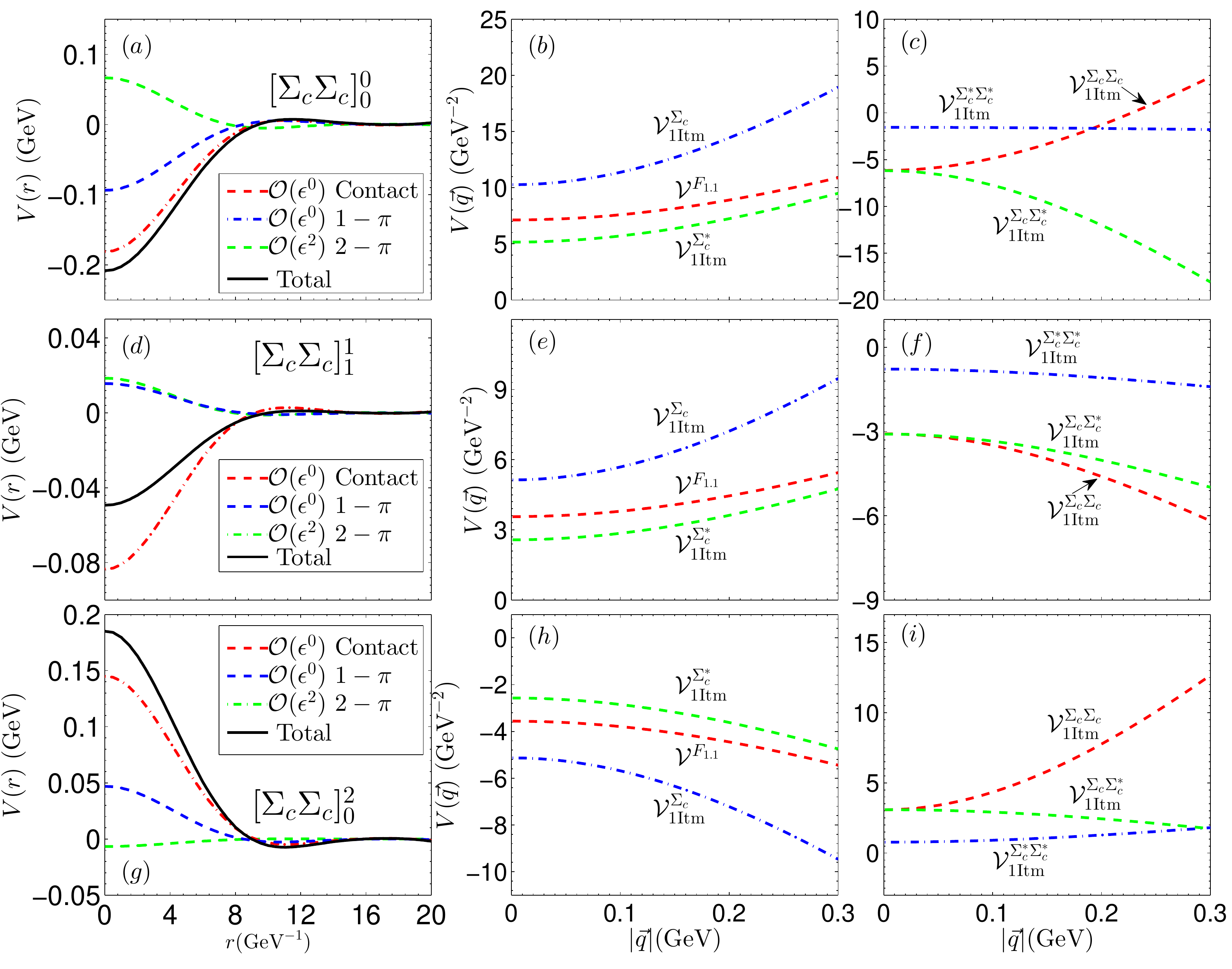}\\
\caption{The first, second, and third rows are the effective
potentials of the $[\Sigma_c\Sigma_c]_0^0$,
$[\Sigma_c\Sigma_c]_1^1$, and $[\Sigma_c\Sigma_c]_0^2$ systems,
respectively. The $\mathcal{V}_{\rm{1Itm}}^{\Sigma^*_c}$,
$\mathcal{V}_{\rm{1Itm}}^{\Sigma_c\Sigma^*_c}$ and
$\mathcal{V}_{1\rm{Itm}}^{\Sigma^*_c\Sigma^*_c}$ are the results of
the $\mathcal{V}^{T_{1.3}}+\mathcal{V}^{T_{1.6}}$,
$\mathcal{V}^{B_{1.8}}+\mathcal{V}^{B_{1.9}}+\mathcal{V}^{R_{1.8}}+\mathcal{V}^{R_{1.9}}$,
and $\mathcal{V}^{B_{1.7}}+\mathcal{V}^{R_{1.7}}$ in momentum space,
respectively. The other notations are the same as in Fig.
\ref{sigcsigccase1}.\label{sigcsigccase2}}
\end{figure*}

In the heavy quark limit, the $\Sigma_c$ and $\Sigma_c^*$ baryons
are degenerate. In order to evaluate the contributions of the TPE
diagrams, we divide the Case-II into two scenarios---without and
with considering the mass difference $\delta_a$ between the
intermediate $\Sigma_c^*$ and initial $\Sigma_c$ baryon. We present
the effective potentials with $\delta_a=0$ for the
$[\Sigma_c\Sigma_c]_0^{0,2}$ and $[\Sigma_c\Sigma_c]_1^1$ systems in
Fig. \ref{sigcsigccase2}. In
Figs.~\ref{sigcsigccase2}(a),~\ref{sigcsigccase2}(d),
and~\ref{sigcsigccase2}(g),
the inclusion of $\Sigma_c^*$-related channels gives minor corrections
to the TPE potentials of the $\Sigma_c\Sigma_c$ systems. The contributions
of the triangle diagrams with the $\Sigma_c^*$ as the intermediate
state is slightly smaller than that of triangle diagrams with
the $\Sigma_c$ as the intermediate state, as can be seen from
Figs.~\ref{sigcsigccase2}(b),~\ref{sigcsigccase2}(e),
and~\ref{sigcsigccase2}(h). The contributions of the box plus
cross diagrams with the $\Sigma_c\Sigma_c$ and $\Sigma_c\Sigma_c^*$
as the intermediate channels are comparable to each other, and
are bigger than that of the box plus cross diagrams with
the $\Sigma_c^*\Sigma_c^*$ as the intermediate channels.

Then we discuss the scenario of considering the mass splitting
$\delta_a$ in the TPE diagrams. In this scenario, we still have the
relation
\begin{eqnarray}
|\mathcal{V}_{1\rm{Itm}}^{\Sigma_c}|>|\mathcal{V}_{1\rm{Itm}}^{\Sigma_c^*}|
\end{eqnarray}
for the triangle diagrams. However, for the box plus cross diagrams,
we obtain the following relation
\begin{eqnarray}
|\mathcal{V}_{1\rm{Itm}}^{\Sigma_c\Sigma_c}|\approx
|\mathcal{V}_{1\rm{Itm}}^{\Sigma_c^*\Sigma_c^*}|<|\mathcal{V}_{1\rm{Itm}}^{\Sigma_c\Sigma_c^*}|,
\end{eqnarray}
i.e., the contributions of the box plus cross diagrams with the
$\Sigma^*_c\Sigma_c^*$ as the intermediate state are comparable to
that of the box plus cross diagrams with the $\Sigma_c\Sigma_c$ as
intermediate state and are much smaller than that of the box plus
cross diagrams with the $\Sigma_c\Sigma_c^*$ as the intermediate
state. In this scenario, the $[\Sigma_c\Sigma_c]^{0,2}_0$ and
$[\Sigma_c\Sigma_c]_1^1$ are all deeply bound (the binding energies
are all larger than $100$ MeV). The main reason of the unnaturally
large binding energies is that we used the dimensional regulation
scheme to calculate the TPE loop diagrams. The chiral loops
calculated with this regulation scheme will lead to convergence
problems~\cite{Borasoy:1998uu,Donoghue:1998rp,Donoghue:1998bs}. The
results of including the mass splittings might be uncontrollable in
the present dimensional regulation scheme. Thus, in the following,
we only present the results without the mass differences in the TPE
potentials.

\begin{figure*}[htbp]
\includegraphics[width=0.85\linewidth]{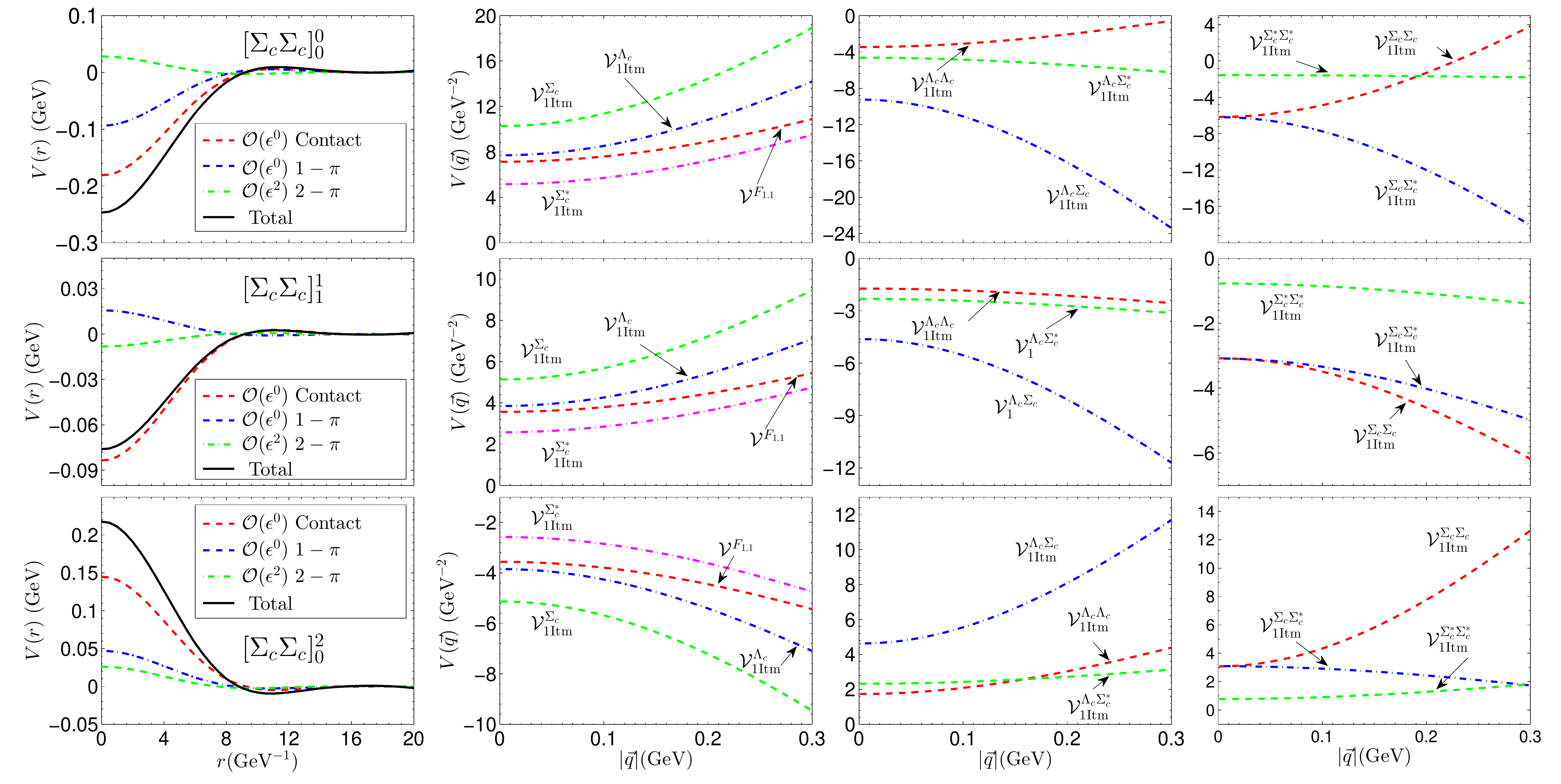}\\
\caption{The first, second, and third rows are the effective
potentials of the $[\Sigma_c\Sigma_c]_0^0$,
$[\Sigma_c\Sigma_c]_1^1$, and $[\Sigma_c\Sigma_c]_0^2$ systems,
respectively. The $\mathcal{V}_{\rm{1Itm}}^{\Lambda_c}$,
$\mathcal{V}_{\rm{1Itm}}^{\Lambda_c\Lambda_c}$,
$\mathcal{V}_{1\rm{Itm}}^{\Lambda_c\Sigma_c}$, and
$\mathcal{V}_{1\rm{Itm}}^{\Lambda_c\Sigma^*_c}$ are the results of
the $\mathcal{V}^{T_{1.1}}+\mathcal{V}^{T_{1.4}}$,
$\mathcal{V}^{B_{1.1}}+\mathcal{V}^{R_{1.1}}$,
$\mathcal{V}^{B_{1.2}}+\mathcal{V}^{R_{1.2}}+\mathcal{V}^{B_{1.3}}+\mathcal{V}^{R_{1.3}}$,
and
$\mathcal{V}^{B_{1.4}}+\mathcal{V}^{R_{1.4}}+\mathcal{V}^{B_{1.5}}+\mathcal{V}^{R_{1.5}}$
in momentum space, respectively. The other notations are the same as
in Fig. \ref{sigcsigccase1} and Fig.
\ref{sigcsigccase2}.\label{sigcsigcpot}}
\end{figure*}

The mass difference $\delta_b$ ($\delta_c$) between the $\Lambda_c$
and $\Sigma_c^{(\ast)}$ baryons are from that of the scalar and
vector diquarks inside the $\Lambda_c$ and $\Sigma_c^{(\ast)}$
baryons. However, if we take these mass differences into account, we
have to face the above mentioned convergence problem again. To
roughly estimate the effects of including the $\Lambda_c$-related
channels, we will not consider the mass differences $\delta_b$ and
$\delta_c$ in the TPE potentials. The results in Case-III for the
$[\Sigma_c\Sigma_c]_0^{0,2}$, $[\Sigma_c\Sigma_c]_1^1$ systems are
presented in Fig.~\ref{sigcsigcpot}. One sees that including the
$\Lambda_c$-related intermediate channels can give very important
corrections to the TPE potentials, and the relative contributions of
the intermediate channels roughly have the following relations
\begin{eqnarray}
|\mathcal{V}_{1\rm{Itm}}^{\Sigma_c}|&>&|\mathcal{V}_{1\rm{Itm}}^{\Lambda_c}|>|\mathcal{V}_{1\rm{Itm}}^{\Sigma^*_c}|,\\
|\mathcal{V}_{1\rm{Itm}}^{\Lambda_c\Sigma_c}|&>&|\mathcal{V}_{1\rm{Itm}}^{\Lambda_c\Lambda_c}|\approx|\mathcal{V}_{1\rm{Itm}}^{\Lambda_c\Sigma^*_c}|,\\
|\mathcal{V}_{1\rm{Itm}}^{\Sigma_c\Sigma_c}|&\approx&|\mathcal{V}_{1\rm{Itm}}^{\Sigma_c\Sigma^*_c}|>|\mathcal{V}_{1\rm{Itm}}^{\Sigma^*_c\Sigma^*_c}|.
\end{eqnarray}
The above relations are consistent with the general understanding of
the couple-channel effect.

The total effective potentials of the $[\Sigma_c\Sigma_c]_0^0$ and
$[\Sigma_c\Sigma_c]_1^1$ are both attractive and can form bound
states. The corresponding binding energies are $101.9$ MeV and $6.8$
MeV, respectively. Since the width of the $\Sigma_c^+$ baryon
is about $2$ MeV and its dominant decay mode is $\Sigma^+_c\rightarrow \Lambda^+_c\pi^0$~\cite{ParticleDataGroup:2020ssz}. Thus, these two bound states are expected to be narrow. One may find the $[\Sigma_c\Sigma_c]_0^0$ bound state in the $\Lambda_c\Lambda_c$ invariant mass distributions. 
For the shallow bound state $[\Sigma_c\Sigma_c]_1^1$, it can be
detected in the $\Lambda_c\pi\Lambda_c\pi$ and
$\Lambda_c\Lambda_c\pi$ final states.

With the same framework, we further study the interactions of the
$\Lambda_c\Lambda_c$ and $\Lambda_c\Sigma_c$ systems. The symmetry
allowed systems are $[\Lambda_c\Lambda_c]_0^0$ and
$[\Lambda_c\Sigma_c]_{0,1}^1$. We present our results for these two
systems in Appendix~\ref{app2}.

\section{Summary}\label{sec4}

In this work, we have studied the interactions of the
$\Sigma_c\Sigma_c$ systems within $\chi$EFT. We introduce the
contact, OPE, and TPE interactions for the
$[\Sigma_c\Sigma_c]_0^{0,2}$ and $[\Sigma_c\Sigma_c]_1^1$ systems
and determine the LECs from the $\Sigma_c^{(*)}\bar{D}^{(*)}$
systems via a quark-level interaction.

We explore the effects of different intermediate channels in the TPE
diagrams in three cases. We introduce the (i) $\Sigma_c$,  (ii)
$\Sigma_c$, $\Sigma_c^{*}$  and  (iii) $\Lambda_c$, $\Sigma_c$,
$\Sigma_c^*$ as the possible intermediate states gradually. We find
the convergence of the chiral expansion is not good if the mass
splittings are explicitly considered in the TPE potentials within
the dimensional regularization scheme, which results in deeply bound
molecular states with unnaturally large binding energies. To cure
this problem, we neglect the mass differences between the initial
$\Sigma_c$ and intermediate ($\Lambda_c$, $\Sigma_c$, $\Sigma_c^*$)
baryons. In this case, the interactions of the
$[\Sigma_c\Sigma_c]_0^0$ and $[\Sigma_c\Sigma_c]_1^1$ are
attractive, while the interaction of the $[\Sigma_c\Sigma_c]_0^2$
system is repulsive. Among the three cases, the TPE potentials in
Case-II are very close to that of Case-I, and the NLO TPE potentials
are comparable to the LO OPE potentials. But if we include the
$\Lambda_c$ as the intermediate channels, the TPE potentials for the
$[\Sigma_c\Sigma_c]_1^1$ and $[\Sigma_c\Sigma_c]_0^2$ change their
signs and the power counting works well.

We obtain two bound states in the $\Sigma_c\Sigma_c$ systems---the
$[\Sigma_c\Sigma_c]_0^0$ and $[\Sigma_c\Sigma_c]_1^1$. These two
states should be narrow ones due to the small width of the
$\Sigma_c$ baryon. The $[\Sigma_c\Sigma_c]_0^0$ and
$[\Sigma_c\Sigma_c]_1^1$ may be reconstructed in the
$\Lambda_c\Lambda_c$ and ($\Lambda^+_c\pi\Lambda_c^+\pi$,
$\Lambda_c^+\Lambda_c^+\pi$) final states, respectively. We hope
that the experiments at LHC, J-PARC, and RHIC can search for these
two states in the future.

\section*{Acknowledgments}
This work is supported by the
National Natural Science Foundation of China under Grants No.
11975033, No. 12105072 and No. 12070131001, the Youth Funds of Hebei
Province (No. A2021201027) and the Start-up Funds for Young Talents
of Hebei University (No. 521100221021).

\begin{appendix}

\section{The interactions of the $\Lambda_c\Lambda_c$ and $\Lambda_c\Sigma_c$ systems}\label{app2}

\subsection{The effective potentials for the $\Lambda_c\Lambda_c$ and $\Lambda_c\Sigma_c$ systems}

Since the vertex $\Lambda_c\Lambda_c\pi$ does not exist, there is no
OPE contribution in the $\Lambda_c\Lambda_c$ system. The contact
potential reads
\begin{eqnarray}
\mathcal{V}_{\Lambda_c\Lambda_c}^{\mathrm{ct}}=-16C_a.\label{LLCon}
\end{eqnarray}
Because the $\Lambda_c$ baryon has a spin-$0$ diquark, the spin-spin
interaction from the light degree of freedom vanishes and only the
central interaction term survives.

For the $\Lambda_c\Sigma_c$ system, the corresponding contact and
OPE potentials read
\begin{eqnarray}
\mathcal{V}_{\Lambda_c\Sigma_c}^{\mathrm{ct}}&=&2C_b,\label{LSCon}\\
\mathcal{V}_{\Lambda_c\Sigma_c}^{\mathrm{OPE}}&=&-\frac{g_2^2}{2f_\pi^2}\frac{\left(\bm{\sigma}_1\cdot\bm{q}\right)\left(\bm{\sigma}_2\cdot\bm{q}\right)}{\bm{q}^2+m_\pi^2}.\label{LSOpe}
\end{eqnarray}
From Eq. (\ref{LSCon}), the contact potentials of the
$[\Lambda_c\Sigma_c]^1_{0}$ and $[\Lambda_c\Sigma_c]_1^1$ states are
the same. The differences between the effective potentials of these
two states will be manifested in their OPE and TPE potentials.

\begin{figure*}[htbp]
\includegraphics[width=0.8\linewidth]{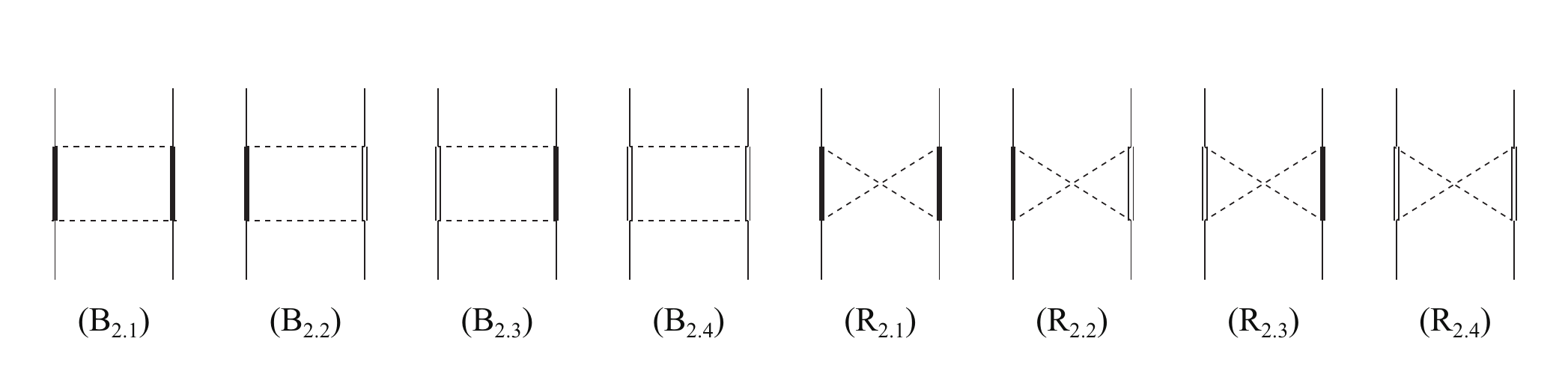}\\
\caption{Two-pion-exchange diagrams that account for the effective
potentials of the $\Lambda_c\Lambda_c$ system at next-to-leading
order. These diagrams include the box diagram ($B_{2.i}$) and cross
diagram ($R_{2.i}$). We use the thin line, thick line, double-thin
line, and dashed line to denote the $\Lambda_c$, $\Sigma_c$,
$\Sigma_c^*$, and $\pi$, respectively.}\label{lamclamctwopionfig}
\end{figure*}

\begin{figure*}[htbp]
\includegraphics[width=0.8\linewidth]{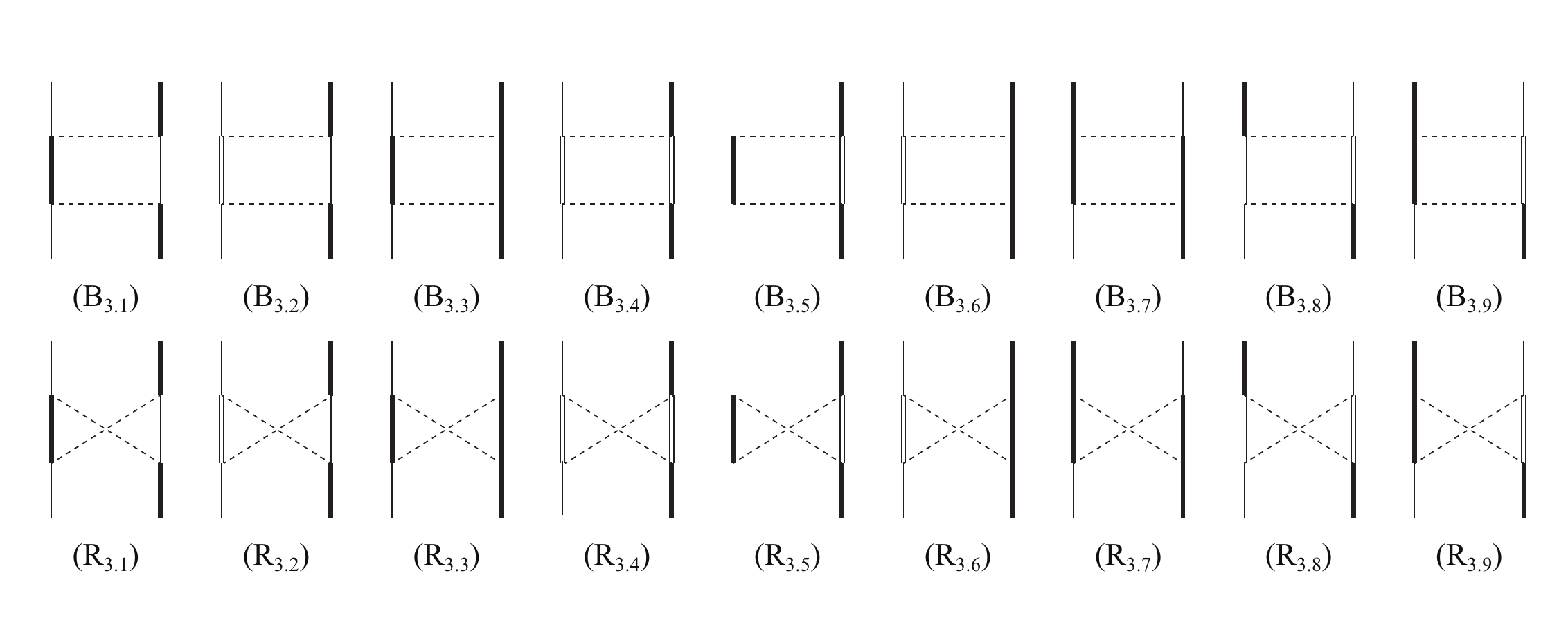}\\
\caption{Two-pion-exchange diagrams for the $\Lambda_c\Sigma_c$
system. The notations are the same as those in Fig.
\ref{lamclamctwopionfig}}\label{lamcsigctwopionfig}
\end{figure*}

The TPE diagrams for the $\Lambda_c\Lambda_c$ and
$\Lambda_c\Sigma_c$ systems are presented in Fig.
\ref{lamclamctwopionfig} and Fig. \ref{lamcsigctwopionfig},
respectively. Since the $\Lambda_c\Lambda_c\pi\pi$ vertex does not
exist, the $\Lambda_c\Lambda_c$ system does not have the football or
triangle diagrams. Since the $\Lambda_c\Lambda_c\pi$ vertex is
forbidden, as presented in Fig. \ref{lamclamctwopionfig}, the
intermediate channels for the $\Lambda_c\Lambda_c$ system can
only be the $\Sigma_c\Sigma_c$, $\Sigma_c\Sigma_c^*$, and
$\Sigma_c^*\Sigma_c^*$ in the box diagrams $B_{2.1}-B_{2.4}$
and cross diagrams $R_{2.1}-R_{2.4}$.

Similar to the $\Lambda_c\Lambda_c$ system, the $\Lambda_c\Sigma_c$
system does not have the football diagram. The obtained amplitudes
of the triangle diagrams for the $\Lambda_c\Sigma_c$ system vanish.
Thus, we do not depict them in Fig. \ref{lamcsigctwopionfig}. The
box diagrams $B_{3.1}$-$B_{3.9}$ and cross diagrams
${R}_{3.1}$-$R_{3.9}$ with the final states unchanged
$(B/R)_{3.1}$-$(B/R)_{3.6}$ and
interchanged
$(B/R)_{3.7}$-$(B/R)_{3.9}$ are
depicted in Fig. \ref{lamcsigctwopionfig}.

The expressions of the box and cross diagrams for the
$\Lambda_c\Lambda_c$ and $\Lambda_c\Sigma_c$ systems can also be
expressed as Eq. (\ref{Bij}) and Eq. (\ref{Rij}), respectively. Note
that the $\Lambda_c\Lambda_c$ and $\Lambda_c\Sigma_c$ can only
couple to the isospin 0 and 1 states, respectively. Thus, the
isospin factors $F_{I}^{\mathrm{B}_{i.j}}$
($F_{I}^{\mathrm{R}_{i.j}}$) defined in Eq. (\ref{Bij}) (Eq.
(\ref{Rij})) are just 1 for each of the box (cross) diagrams in the
$\Lambda_c\Lambda_c$ and $\Lambda_c\Sigma_c$ systems. The other
coefficients defined in Eq. (\ref{Bij}) and Eq. (\ref{Rij}) are
collected in Table \ref{BR factor 2}.

\begin{table*}[!htbp]
\centering
\renewcommand\arraystretch{1.5}
\caption{The coefficients of the two-pion-exchange box and cross
diagrams defined in Eq. (\ref{Bij}) and Eq. (\ref{Rij}) for the
$\Lambda_c\Lambda_c$ and $\Lambda_c\Sigma_c$ systems. \label{BR
factor 2}} \setlength{\tabcolsep}{2.8mm} {
\begin{tabular}{cl|ccccccccccccc}
\toprule[0.7pt]
&&$\mathcal{C}^{(\mathrm{B}/\mathrm{R})_{i.j}}$&$C_1^{(\mathrm{B}/\mathrm{R})_{i.j}}$&$C_2^{(\mathrm{B}/\mathrm{R})_{i.j}}$&$C_3^{(\mathrm{B}/\mathrm{R})_{i.j}}$
&$C_4^{(\mathrm{B}/\mathrm{R})_{i.j}}$&$C_5^{(\mathrm{B}/\mathrm{R})_{i.j}}$&$\mathcal{E}_1^{(\mathrm{B}/\mathrm{R})_{i.j}}$&$\mathcal{E}_2^{(\mathrm{B}/\mathrm{R})_{i.j}}$\\
\hline
\multirow{3}*{$i=2$}&$j=1$&$\frac{3g_2^4}{4}$&$-1$&$1/-1$&$1$&$-10$&$15$&$-\delta_b$&$-\delta_b$\\
&$j=2,3$&$\frac{g_2^2g_4^2}{4}$&$-2$&$-1/1$&$2$&$-20$&$30$&$-\delta_b$&$-\delta_c$\\
&$j=4$&$\frac{g_4^4}{12}$&$-4$&$1/-1$&$4$&$-40$&60&$-\delta_c$&$-\delta_c$\\
\hline
\multirow{9}*{$i=3$}&$j=1$&$\frac{g_2^4}{4}$&$-1$&$1/-1$&$1$&$-10$&$15$&$\delta_b$&$-\delta_b$\\
&$j=2$&$\frac{g_2^2g_4^2}{12}$&$-2$&$-1/1$&$2$&$-20$&$30$&$\delta_b$&$-\delta_c$\\
&$j=3$&$\frac{g_1^2g_2^2}{4}$&$-1$&$1/-1$&$1$&$-10$&$15$&0&$-\delta_b$\\
&$j=4$&$\frac{g_3^2g_4^2}{36}$&$-4$&$1/-1$&$4$&$-40$&$60$&$-\delta_a$&$-\delta_c$\\
&$j=5$&$\frac{g_2^2g_3^2}{12}$&$-2$&$-1/1$&$2$&$-20$&$30$&$-\delta_a$&$-\delta_b$\\
&$j=6$&$\frac{g_1^2g_4^2}{12}$&$-2$&$-1/1$&$2$&$-20$&$30$&$0$&$-\delta_c$\\
&$j=7$&$\frac{g_1^2g_2^2}{4}$&$1/-1$&$-1$&$-1/1$&$10$/$-10$&$-15$/$15$&$0$&$-\delta_b$\\
&$j=8$&$\frac{g_3^2g_4^2}{36}$&$4/-4$&$-1$&$-4/4$&$40$/$-40$&$-60$/$60$&$-\delta_a$&$-\delta_c$\\
&$j=9$&$\frac{g_1g_2g_3g_4}{6}$&$2/-2$&$1$&$-2/2$&$20$/$-20$&$-30$/$30$&$-\delta_a$&$-\delta_b$\\
\bottomrule[0.7pt]
\end{tabular}}
\end{table*}

\subsection{Results and discussion}

The leading order contact terms for the $[\Lambda_c\Lambda_c]_0^0$
and $[\Lambda_c\Sigma_c]_{0,1}^1$ systems at quark level
\cite{Chen:2021cfl} are
\begin{eqnarray}
\mathcal{V}_{[\Lambda_c\Lambda_c]_0^0}&=&\frac{4}{3}\tilde{g}_s,\nonumber\\
\mathcal{V}_{[\Lambda_c\Sigma_c]_{0,1}^1}&=&\frac{4}{3}\tilde{g}_s\nonumber.
\end{eqnarray}
Correspondingly, we obtain the LECs defined in Eq. \ref{FBCI} as
\begin{eqnarray}
C_a&=&-1.3 \quad{\rm GeV}^{-2},\quad C_b=10.4\quad{\rm GeV}^{-2}.
\end{eqnarray}
With the above preparation, we calculate the total effective
potentials for the $[\Lambda_c\Lambda_c]_0^0$ and
$[\Lambda_c\Sigma_c]_{0,1}^1$ systems. Then we solve the
corresponding Schr\"odinger equations to search for the binding
solutions.

We first present the results of the $\Lambda_c\Lambda_c$ system. To
calculate the two-pion-exchange effective potential, we firstly
neglect the mass differences between the intermediate
$\Sigma_c^{(*)}\Sigma_c^{(*)}$ channels with the initial
$\Lambda_c\Lambda_c$ state, i.e., we set $\delta_b=\delta_c=0$.
Correspondingly, we need to adopt
\begin{eqnarray}
J_{ab}^B\left(0,0\right)&=&\frac{\partial}{\partial{x}}J_{ab}^T\left(x\right)\Big|_{x\rightarrow 0},\\
J_{ab}^R\left(0,0\right)&=&-\frac{\partial}{\partial{x}}J_{ab}^T\left(x\right)\Big|_{x\rightarrow0}
\end{eqnarray}
to replace the scalar loop functions in Eq. (\ref{Bij}-\ref{Rij}).

\begin{figure*}[htbp]
\includegraphics[width=0.8\linewidth]{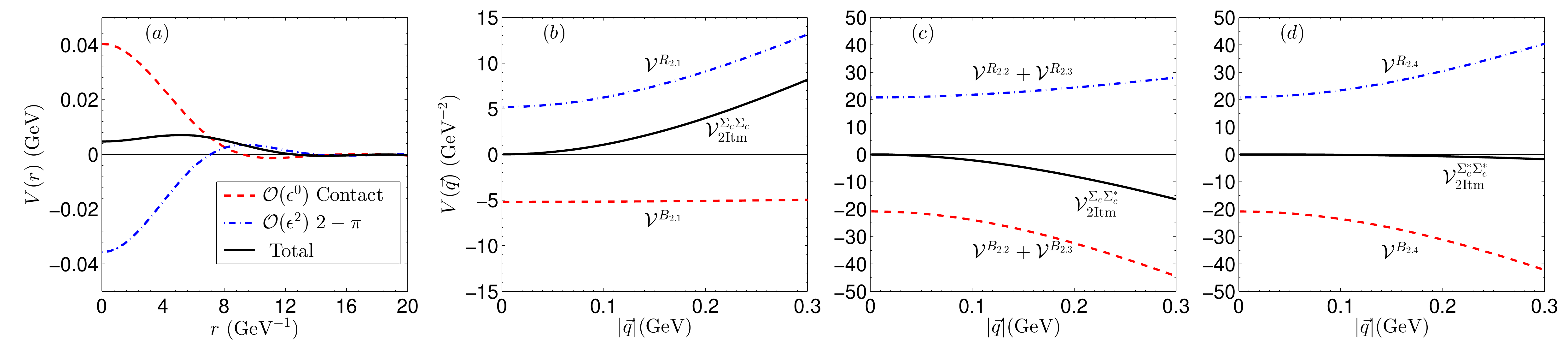}\\
\caption{$(a)$ present the contact, TPE, and total effective
potentials of the $[\Lambda_c\Lambda_c]_0^0$ system in the
coordinate space. $b$, $c$, and $d$ depict the contributions of
two-pion-exchange diagrams in momentum space. The
$\mathcal{V}_{2\rm{Itm}}^{\Sigma_c\Sigma_c}$,
$\mathcal{V}_{2\rm{Itm}}^{\Sigma_c\Sigma_c^*}$, and
$\mathcal{V}_{2\rm{Itm}}^{\Sigma_c^*\Sigma_c^*}$ are the results of
$\mathcal{V}^{B_{2.1}}+\mathcal{V}^{R_{2.1}}$,
$\mathcal{V}^{B_{2.2}}+\mathcal{V}^{B_{2.3}}+\mathcal{V}^{R_{2.2}}+\mathcal{V}^{R_{2.3}}$,
and $\mathcal{V}^{B_{2.4}}+\mathcal{V}^{R_{2.4}}$ that account for
the contributions induced from including the $\Sigma_c\Sigma_c$,
$\Sigma_c\Sigma_c^*$, and $\Sigma_c^*\Sigma_c^*$ intermediate
channels, respectively.\label{lamclamcpot}}
\end{figure*}

In Fig. \ref{lamclamcpot} (a), we present the contact,
two-pion-exchange, and total effective potentials of the
$[\Lambda_c\Lambda_c]_0^0$ systems in coordinate space. As
illustrated in Fig. \ref{lamclamcpot} (a), the contact and
two-pion-exchange potentials provide the repulsive and positive
forces, respectively. Besides, the contact potential is comparable
to the two-pion-exchange potential but with the opposite sign. Note
that the determined contact interaction is indeed a small repulsive
force due to the weak couplings between the two $\Lambda_c$ baryons
with spin-0 diquarks. Thus, the total effective potential of the
$[\Lambda_c\Lambda_c]_0^0$ system is very weak. Of course, the
obtained potential can not form a $[\Lambda_c\Lambda_c]_0^0$ bound
state.

Since the magnitude of the contributions from the two-pion-exchange
diagrams is comparable to that of the contact term, we further check
the relative contributions of the intermediate
$\Sigma_c^{(*)}\Sigma_c^{(*)}$ channels in the $\Lambda_c\Lambda_c$
system. In momentum space, we present the effective potentials of
the $(B_{2.1}, R_{2.1})$, $(B_{2.2}+B_{2.3},R_{2.2}+R_{2.3})$, and
$(B_{2.4},R_{2.4})$ diagrams in Fig. \ref{lamclamcpot} (b), (c), and
(d), respectively. From Fig. \ref{lamclamcpot} (b), we can easily
find out that the $B_{2.1}$ and $R_{2.1}$ provide the attractive and
repulsive forces, respectively. We also sum the
$\mathcal{V}^{B_{2.1}}$ and $\mathcal{V}^{R_{2.1}}$ to give the
total potential induced from including the $\Sigma_c\Sigma_c$
intermediate channel. Summing the $V^{B_{2.2}}+V^{B_{2.3}}$ and
$V^{R_{2.2}}+V^{R_{2.3}}$, we find that the inclusion of the
$\Sigma_c\Sigma_c^*$ channel provides a repulsive force to the
$\Lambda_c\Lambda_c$ system, and this channel is more important than
the $\Sigma_c\Sigma_c$ intermediate channel. Besides, as can be seen
from Fig. \ref{lamclamcpot} (d), the $\Sigma^*_c\Sigma^*_c$ channel
gives very tiny contribution to the two-pion-exchange potential.
Although the contributions of the $V^{B_{2.4}}$ and $V^{R_{2.4}}$
are considerable, their opposite signs make the
$V_{\rm{2Itm}}^{\Sigma_c^*\Sigma_c^*}$ negligible.

We further check the results of the $[\Lambda_c\Lambda_c]_0^0$ by
including the mass differences between the initial $\Lambda_c$ and
the intermediate $\Sigma_c/\Sigma_c^{*}$ states. However, we find
that the convergence of the chiral series problem also exists in the
$\Lambda_c\Lambda_c$ system. After considering the mass differences,
the contributions from the two-pion-exchange diagrams are
significantly magnified. The obtained two-pion-exchange potential is
much larger than the leading order contact potential, which may
violate the power counting rule. The unexpected total large
potential will lead to a very deeply bound state with the binding
energy 197 MeV.

We also check the relative contributions from the intermediate
$\Sigma_c\Sigma_c$, $\Sigma_c\Sigma_c^*$, and $\Sigma_c^*\Sigma_c^*$
channels. After we include the mass differences, these three
channels all provide strong attractive forces and have
\begin{eqnarray}
\left|\mathcal{V}_{1\rm{Itm}}^{\Sigma_c\Sigma_c}\right|<\left|\mathcal{V}_{1\rm{Itm}}^{\Sigma_c\Sigma^*_c}\right|<\left|\mathcal{V}_{1\rm{Itm}}^{\Sigma^*_c\Sigma^*_c}\right|.
\end{eqnarray}
Generally, the intermediate channel would have less influence to the
studied two-body system if its threshold lies further away from the
threshold of the considered system. The obtained results contradicts
this argument. The $[\Lambda_c\Sigma_c]_{0,1}^1$ have very similar
results due to the uncertainties introduced from the nonanalytic
chiral loops in the two-pion-exchange loop diagrams. Thus, in the
following, we only discuss the $\Lambda_c\Sigma_c$ system without
considering the mass differences between the initial and
intermediate baryons, i.e., we adopt $\delta_a=\delta_b=\delta_c$.

\begin{figure*}
\includegraphics[width=0.7\linewidth]{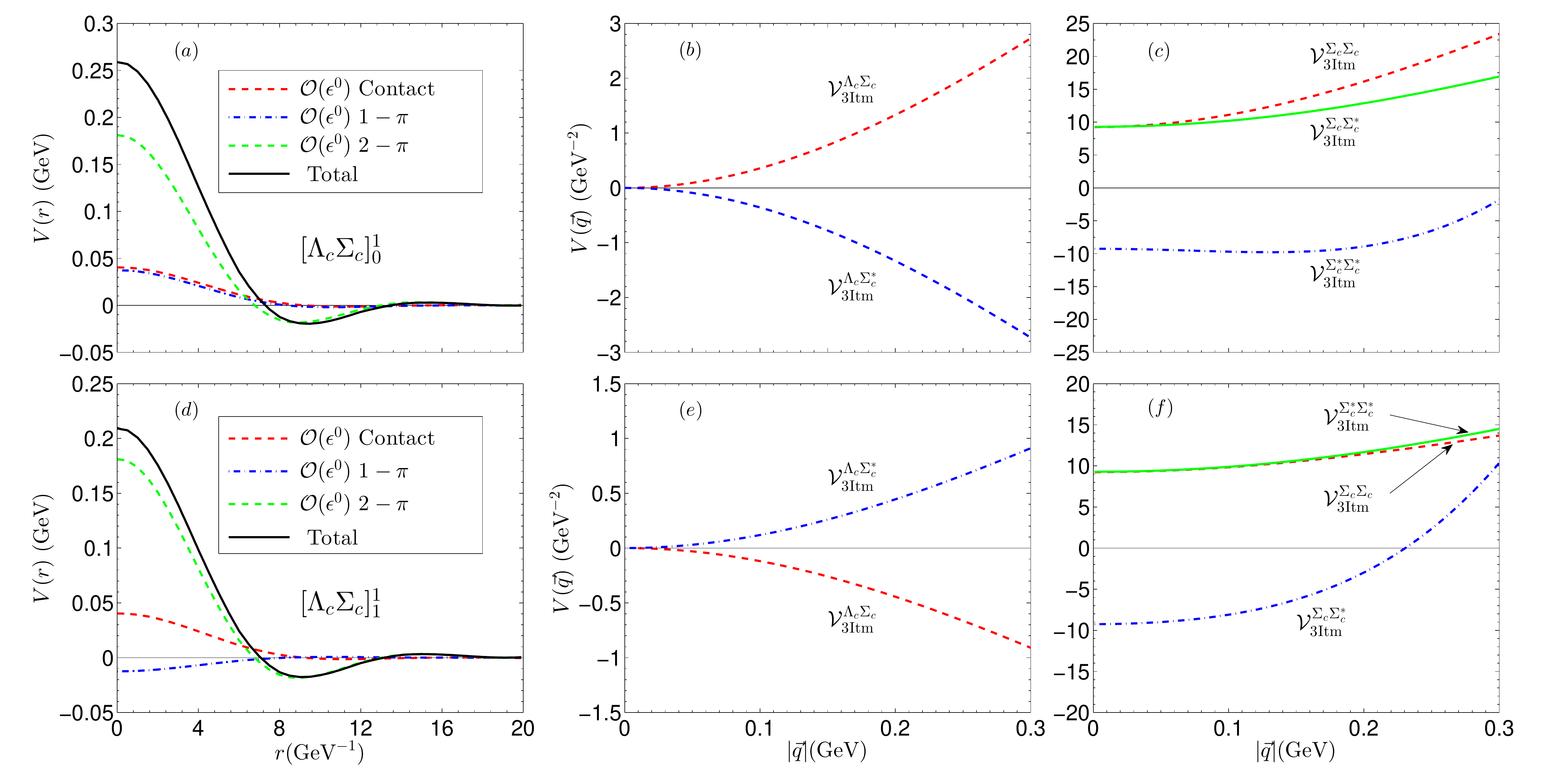}
\caption{$(a)$ and $(d)$ present the contact, OPE, TPE, and total
effective potentials of the $[\Lambda_c\Sigma_c]_0^0$ system in the
coordinate space. $(b)$, $(c)$, $(e)$ $(f)$ depict the contributions
of two-pion-exchange diagrams. The
$\mathcal{V}_{3\rm{Itm}}^{\Lambda_c\Sigma_c}$,
$\mathcal{V}_{3\rm{Itm}}^{\Lambda_c\Sigma_c^*}$,
$\mathcal{V}_{3\rm{Itm}}^{\Sigma_c\Sigma_c}$,
$\mathcal{V}_{3\rm{Itm}}^{\Sigma_c\Sigma_c^*}$, and
$\mathcal{V}_{3\rm{Itm}}^{\Sigma^*_c\Sigma^*_c}$ are the results of
$\mathcal{V}^{B_{3.1}}+\mathcal{V}^{R_{3.1}}$,
$\mathcal{V}^{B_{3.2}}+\mathcal{V}^{R_{3.2}}$,
$\mathcal{V}^{B_{3.3}}+\mathcal{V}^{R_{3.3}}+\mathcal{V}^{B_{3.7}}+\mathcal{V}^{R_{3.7}}$,
$\mathcal{V}^{B_{3.5}}+\mathcal{V}^{R_{3.5}}+\mathcal{V}^{B_{3.6}}+\mathcal{V}^{R_{3.6}}+\mathcal{V}^{B_{3.9}}+\mathcal{V}^{R_{3.9}}$,
and
$\mathcal{V}^{B_{3.4}}+\mathcal{V}^{R_{3.4}}+\mathcal{V}^{B_{3.8}}+\mathcal{V}^{R_{3.8}}$
that account for the contributions induced from including the
$\Lambda_c\Sigma_c$, $\Lambda_c\Sigma_c^*$, $\Sigma_c\Sigma_c$,
$\Sigma_c\Sigma_c^*$, and $\Sigma_c^*\Sigma_c^*$ intermediate
channels, respectively.\label{lamcsigcpot}}
\end{figure*}

The effective potentials of the $[\Lambda_c\Sigma_c]_{0,1}^1$ system
are presented in Fig. \ref{lamcsigcpot}. The leading order contact
potentials of the $[\Lambda_c\Sigma_c]_{0,1}^1$ systems arise from
the interactions of their light degrees of freedom (d.o.f). The
$[\Lambda_c\Sigma_c]_{0}^1$ and $[\Lambda_c\Sigma_c]_1^1$ have the
same contact potentials. The corrections from the spin of heavy
quarks are manifest after we include the one-pion-exchange and
two-pion-exchange interactions. As presented in Fig.
\ref{lamcsigcpot} (a) and (d), the one-pion-exchange potentials
provide the repulsive and attractive forces in the
$[\Lambda_c\Sigma_c]_0^1$ and $[\Lambda_c\Sigma_c]_1^1$ systems,
respectively. In both systems, the magnitudes of the
one-pion-exchange and contact contributions are comparable to each
other and much smaller than those of the two-pion-exchange
potentials. Note that the contact interaction for the
$\Lambda_c\Sigma_c$ system determined from quark model is identical
to that of the $\Lambda_c\Lambda_c$ system. Thus, the contact
interactions of the $[\Lambda_c\Sigma_c]_{0,1}^1$ systems are
relatively small. For such a system, it is very likely that their
two-pion-exchange potentials are larger than its contact term.

We further check the relative contributions from different
intermediate channels. We present the effective potentials induced
from the intermediate $\Lambda_c\Sigma_c^{(*)}$ and
$\Sigma_c^{(*)}\Sigma_c^{(*)}$ channels in momentum space in Fig.
\ref{lamcsigcpot} (b), (c), (e), and (f). The $\Lambda_c\Sigma_c$
and $\Lambda_c\Sigma_c^*$ have relatively small contributions to the
two-pion-exchange potential, while the contributions of the
$\Sigma_c\Sigma_c$, $\Sigma_c\Sigma_c^*$ and $\Sigma^*_c\Sigma^*_c$
channels are comparable to each other and quite important.

\end{appendix}


\begin{thebibliography}{199}
\bibitem{Godfrey:1985xj}
S.~Godfrey and N.~Isgur,
\href{https://journals.aps.org/prd/abstract/10.1103/PhysRevD.32.189}{Phys.
Rev. D \textbf{32} (1985), 189-231}
\bibitem{Capstick:1985xss}
S.~Capstick and N.~Isgur,
\href{https://aip.scitation.org/doi/abs/10.1063/1.35361}{AIP Conf.
Proc. \textbf{132} (1985), 267-271}
\bibitem{BaBar:2003oey}
B.~Aubert \textit{et al.} [BaBar],
\href{https://journals.aps.org/prl/abstract/10.1103/PhysRevLett.90.242001}{Phys.
Rev. Lett. \textbf{90} (2003), 242001}
\bibitem{Belle:2003nnu}
S.~K.~Choi \textit{et al.} [Belle],
\href{https://journals.aps.org/prl/abstract/10.1103/PhysRevLett.91.262001}{Phys.
Rev. Lett. \textbf{91} (2003), 262001}
\bibitem{ParticleDataGroup:2020ssz}
P.~A.~Zyla \textit{et al.} [Particle Data Group],
\href{https://academic.oup.com/ptep/article/doi/10.1093/ptep/ptaa104/5891211}{PTEP
\textbf{2020}, no.8, 083C01 (2020).}

\bibitem{Chen:2016qju}
H.~X.~Chen, W.~Chen, X.~Liu and S.~L.~Zhu,
\href{https://www.sciencedirect.com/science/article/pii/S037015731630103X?via%3Dihub}{Phys. Rept. \textbf{639}, 1-121 (2016)}
\bibitem{Liu:2019zoy}
Y.~R.~Liu, H.~X.~Chen, W.~Chen, X.~Liu and S.~L.~Zhu,
\href{https://www.sciencedirect.com/science/article/pii/S0146641019300304?via%3Dihub}{Prog. Part. Nucl. Phys. \textbf{107}, 237-320 (2019)}
\bibitem{Liu:2013waa}
X.~Liu,
\href{https://link.springer.com/article/10.1007%2Fs11434-014-0407-2}{Chin. Sci. Bull. \textbf{59}, 3815-3830 (2014)}
\bibitem{Guo:2017jvc}
F.~K.~Guo, C.~Hanhart, U.~G.~Mei\ss{}ner, Q.~Wang, Q.~Zhao and
B.~S.~Zou,
\href{https://journals.aps.org/rmp/abstract/10.1103/RevModPhys.90.015004}{Rev.
Mod. Phys. \textbf{90}, no.1, 015004 (2018)}
\bibitem{Hosaka:2016pey}
A.~Hosaka, T.~Iijima, K.~Miyabayashi, Y.~Sakai and S.~Yasui,
\href{https://inspirehep.net/literature/1436498}{PTEP \textbf{2016},
no.6, 062C01 (2016)}
\bibitem{Lebed:2016hpi}
R.~F.~Lebed, R.~E.~Mitchell and E.~S.~Swanson,
\href{https://www.sciencedirect.com/science/article/pii/S0146641016300734?via%3Dihub}{Prog. Part. Nucl. Phys. \textbf{93}, 143-194 (2017)}
\bibitem{Esposito:2016noz}
A.~Esposito, A.~Pilloni and A.~D.~Polosa,
\href{https://www.sciencedirect.com/science/article/pii/S037015731630391X?via%3Dihub}{Phys. Rept. \textbf{668}, 1-97 (2017)}
\bibitem{Brambilla:2019esw}
N.~Brambilla, S.~Eidelman, C.~Hanhart, A.~Nefediev, C.~P.~Shen,
C.~E.~Thomas, A.~Vairo and C.~Z.~Yuan,
\href{https://www.sciencedirect.com/science/article/pii/S0370157320301915?via%3Dihub}{Phys. Rept. \textbf{873}, 1-154 (2020)}
\bibitem{Chen:2022asf}
H.~X.~Chen, W.~Chen, X.~Liu, Y.~R.~Liu and S.~L.~Zhu,
\href{https://arxiv.org/abs/2204.02649}{[arXiv:2204.02649
[hep-ph]].}
\bibitem{Meng:2022ozq}
L.~Meng, B.~Wang, G.~J.~Wang and S.~L.~Zhu,
\href{https://arxiv.org/abs/2204.08716}{arXiv:2204.08716 [hep-ph]}.
\bibitem{Aaij:2015tga}
R.~Aaij {\it et al.} [LHCb Collaboration],
\href{https://journals.aps.org/prl/abstract/10.1103/PhysRevLett.115.072001}{Phys.\
Rev.\ Lett.\  {\bf 115}, 072001 (2015).}
\bibitem{Aaij:2016phn}
R.~Aaij {\it et al.} [LHCb Collaboration],
\href{https://journals.aps.org/prl/abstract/10.1103/PhysRevLett.117.082002}{Phys.\
Rev.\ Lett.\  {\bf 117}, no. 8, 082002 (2016).}
\bibitem{Aaij:2019vzc}
  R.~Aaij {\it et al.} [LHCb Collaboration],
\href{https://journals.aps.org/prl/abstract/10.1103/PhysRevLett.122.222001}{Phys.\
Rev.\ Lett.\  {\bf 122}, no. 22, 222001 (2019).}
\bibitem{Wu:2010jy}
J.~J.~Wu, R.~Molina, E.~Oset and B.~S.~Zou,
\href{https://journals.aps.org/prl/abstract/10.1103/PhysRevLett.105.232001}{Phys.
Rev. Lett. \textbf{105}, 232001 (2010)}
\bibitem{Yang:2011wz}
Z.~C.~Yang, Z.~F.~Sun, J.~He, X.~Liu and S.~L.~Zhu,
\href{https://iopscience.iop.org/article/10.1088/1674-1137/36/1/002}{Chin.
Phys. C \textbf{36}, 6-13 (2012)}
\bibitem{Wang:2011rga}
W.~L.~Wang, F.~Huang, Z.~Y.~Zhang and B.~S.~Zou,
\href{https://journals.aps.org/prc/abstract/10.1103/PhysRevC.84.015203}{Phys.
Rev. C \textbf{84}, 015203 (2011)}
\bibitem{Chen:2019asm}
R.~Chen, Z.~F.~Sun, X.~Liu and S.~L.~Zhu,
\href{https://journals.aps.org/prd/abstract/10.1103/PhysRevD.100.011502}{Phys.
Rev. D \textbf{100}, no.1, 011502 (2019)}
\bibitem{Liu:2019tjn}
M.~Z.~Liu, Y.~W.~Pan, F.~Z.~Peng, M.~S\'anchez S\'anchez,
L.~S.~Geng, A.~Hosaka and M.~Pavon Valderrama,
\href{https://journals.aps.org/prl/abstract/10.1103/PhysRevLett.122.242001}{Phys.
Rev. Lett. \textbf{122}, no.24, 242001 (2019)}
\bibitem{He:2019ify}
J.~He,
\href{https://link.springer.com/article/10.1140%2Fepjc%2Fs10052-019-6906-1}{Eur. Phys. J. C \textbf{79}, no.5, 393 (2019)}
\bibitem{Xiao:2019aya}
C.~W.~Xiao, J.~Nieves and E.~Oset,
\href{https://journals.aps.org/prd/abstract/10.1103/PhysRevD.100.014021}{Phys.
Rev. D \textbf{100}, no.1, 014021 (2019)}
\bibitem{Meng:2019ilv}
L.~Meng, B.~Wang, G.~J.~Wang and S.~L.~Zhu,
\href{https://journals.aps.org/prd/abstract/10.1103/PhysRevD.100.014031}{Phys.
Rev. D \textbf{100}, no.1, 014031 (2019)}
\bibitem{Yamaguchi:2019seo}
Y.~Yamaguchi, H.~Garc\'\i{}a-Tecocoatzi, A.~Giachino, A.~Hosaka,
E.~Santopinto, S.~Takeuchi and M.~Takizawa,
\href{https://journals.aps.org/prd/abstract/10.1103/PhysRevD.101.091502}{Phys.
Rev. D \textbf{101}, no.9, 091502 (2020)}
\bibitem{PavonValderrama:2019nbk}
M.~Pavon Valderrama,
\href{https://journals.aps.org/prd/abstract/10.1103/PhysRevD.100.094028}{Phys.
Rev. D \textbf{100}, no.9, 094028 (2019)}

\bibitem{Chen:2019bip}
H.~X.~Chen, W.~Chen and S.~L.~Zhu,
\href{https://journals.aps.org/prd/abstract/10.1103/PhysRevD.100.051501}{Phys.
Rev. D \textbf{100}, no.5, 051501 (2019)}
\bibitem{Burns:2019iih}
T.~J.~Burns and E.~S.~Swanson,
\href{https://journals.aps.org/prd/abstract/10.1103/PhysRevD.100.114033}{Phys.
Rev. D \textbf{100}, no.11, 114033 (2019)}
\bibitem{Du:2019pij}
M.~L.~Du, V.~Baru, F.~K.~Guo, C.~Hanhart, U.~G.~Mei\ss{}ner,
J.~A.~Oller and Q.~Wang,
\href{https://journals.aps.org/prl/abstract/10.1103/PhysRevLett.124.072001}{Phys.
Rev. Lett. \textbf{124}, no.7, 072001 (2020)}
\bibitem{Wang:2019ato}
B.~Wang, L.~Meng and S.~L.~Zhu,
\href{https://link.springer.com/article/10.1007%2FJHEP11%282019%29108}{JHEP \textbf{11}, 108 (2019)}
\bibitem{Chen:2016ryt}
R.~Chen, J.~He and X.~Liu,
\href{https://iopscience.iop.org/article/10.1088/1674-1137/41/10/103105}{Chin.
Phys. C \textbf{41}, no.10, 103105 (2017)}
\bibitem{Santopinto:2016pkp}
E.~Santopinto and A.~Giachino,
\href{https://journals.aps.org/prd/abstract/10.1103/PhysRevD.96.014014}{Phys.
Rev. D \textbf{96}, no.1, 014014 (2017)}
\bibitem{Shen:2019evi}
C.~W.~Shen, J.~J.~Wu and B.~S.~Zou,
\href{https://journals.aps.org/prd/abstract/10.1103/PhysRevD.100.056006}{Phys.
Rev. D \textbf{100}, no.5, 056006 (2019)}
\bibitem{Xiao:2019gjd}
C.~W.~Xiao, J.~Nieves and E.~Oset,
\href{https://www.sciencedirect.com/science/article/pii/S0370269319307737?via%3Dihub}{Phys. Lett. B \textbf{799}, 135051 (2019)}
\bibitem{Wang:2019nvm}
B.~Wang, L.~Meng and S.~L.~Zhu,
\href{https://journals.aps.org/prd/abstract/10.1103/PhysRevD.101.034018}{Phys.
Rev. D \textbf{101}, no.3, 034018 (2020)}
\bibitem{Chen:2015sxa}
H.~X.~Chen, L.~S.~Geng, W.~H.~Liang, E.~Oset, E.~Wang and J.~J.~Xie,
\href{https://journals.aps.org/prc/abstract/10.1103/PhysRevC.93.065203}{Phys.
Rev. C \textbf{93}, no.6, 065203 (2016)}
\bibitem{LHCb:2020jpq}
R.~Aaij \textit{et al.} [LHCb],
\href{https://www.sciencedirect.com/science/article/pii/S2095927321001717?via%3Dihub}{Sci. Bull. \textbf{66}, 1278-1287 (2021)}
\bibitem{LHCb:2021chn}
R.~Aaij \textit{et al.} [LHCb],
\href{https://journals.aps.org/prl/abstract/10.1103/PhysRevLett.128.062001}{Phys.
Rev. Lett. \textbf{128} (2022), 062001}
\bibitem{LHCb:2021auc}
R.~Aaij \textit{et al.} [LHCb],
\href{https://arxiv.org/abs/2109.01056}{[arXiv:2109.01056
[hep-ex]].}
\bibitem{LHCb:2021vvq}
R.~Aaij \textit{et al.} [LHCb],
\href{https://arxiv.org/abs/2109.01038}{[arXiv:2109.01038
[hep-ex]].}
\bibitem{Carlson:1987hh}
J.~Carlson, L.~Heller and J.~A.~Tjon,
\href{https://journals.aps.org/prd/abstract/10.1103/PhysRevD.37.744}{Phys.
Rev. D \textbf{37}, 744 (1988)}
\bibitem{Gelman:2002wf}
B.~A.~Gelman and S.~Nussinov,
\href{https://www.sciencedirect.com/science/article/pii/S0370269302030691?via%3Dihub}{Phys. Lett. B \textbf{551}, 296-304 (2003)}
\bibitem{Vijande:2003ki}
J.~Vijande, F.~Fernandez, A.~Valcarce and B.~Silvestre-Brac,
\href{https://link.springer.com/article/10.1140%2Fepja%2Fi2003-10128-9}{Eur. Phys. J. A \textbf{19}, 383 (2004)}
\bibitem{Cui:2006mp}
Y.~Cui, X.~L.~Chen, W.~Z.~Deng and S.~L.~Zhu,
\href{https://arxiv.org/abs/hep-ph/0607226}{HEPNP \textbf{31}, 7-13
(2007)}
\bibitem{Navarra:2007yw}
F.~S.~Navarra, M.~Nielsen and S.~H.~Lee,
Phys. Lett. B \textbf{649}, 166-172 (2007)
\href{https://www.sciencedirect.com/science/article/pii/S0370269307004376?via%3Dihub}{doi:10.1016/j.physletb.2007.04.010}
\bibitem{Vijande:2007rf}
J.~Vijande, E.~Weissman, A.~Valcarce and N.~Barnea,
\href{https://journals.aps.org/prd/abstract/10.1103/PhysRevD.76.094027}{Phys.
Rev. D \textbf{76}, 094027 (2007)}
\bibitem{Ebert:2007rn}
D.~Ebert, R.~N.~Faustov, V.~O.~Galkin and W.~Lucha,
\href{https://journals.aps.org/prd/abstract/10.1103/PhysRevD.76.114015}{Phys.
Rev. D \textbf{76}, 114015 (2007)}
\bibitem{Lee:2009rt}
S.~H.~Lee and S.~Yasui,
\href{https://link.springer.com/article/10.1140%2Fepjc%2Fs10052-009-1140-x}{Eur. Phys. J. C \textbf{64}, 283-295 (2009)}
\bibitem{Wang:2017uld}
Z.~G.~Wang,
\href{https://www.actaphys.uj.edu.pl/index_n.php?I=R&V=49&N=10#1781}{Acta
Phys. Polon. B \textbf{49} (2018), 1781}
\bibitem{Yang:2009zzp}
Y.~Yang, C.~Deng, J.~Ping and T.~Goldman,
\href{https://journals.aps.org/prd/abstract/10.1103/PhysRevD.80.114023}{Phys.
Rev. D \textbf{80}, 114023 (2009)}
\bibitem{Du:2012wp}
M.~L.~Du, W.~Chen, X.~L.~Chen and S.~L.~Zhu,
\href{https://journals.aps.org/prd/abstract/10.1103/PhysRevD.87.014003}{Phys.
Rev. D \textbf{87}, no.1, 014003 (2013)}
\bibitem{Luo:2017eub}
S.~Q.~Luo, K.~Chen, X.~Liu, Y.~R.~Liu and S.~L.~Zhu,
\href{https://link.springer.com/article/10.1140%2Fepjc%2Fs10052-017-5297-4}{Eur. Phys. J. C \textbf{77}, no.10, 709 (2017)}
\bibitem{Karliner:2017qjm}
M.~Karliner and J.~L.~Rosner,
\href{https://journals.aps.org/prl/abstract/10.1103/PhysRevLett.119.202001}{Phys.
Rev. Lett. \textbf{119}, no.20, 202001 (2017)}
\bibitem{Eichten:2017ffp}
E.~J.~Eichten and C.~Quigg,
\href{https://journals.aps.org/prl/abstract/10.1103/PhysRevLett.119.202002}{Phys.
Rev. Lett. \textbf{119}, no.20, 202002 (2017)}
\bibitem{Sakai:2017avl}
S.~Sakai, L.~Roca and E.~Oset,
\href{https://journals.aps.org/prd/abstract/10.1103/PhysRevD.96.054023}{Phys.
Rev. D \textbf{96}, no.5, 054023 (2017)}
\bibitem{Manohar:1992nd}
A.~V.~Manohar and M.~B.~Wise,
\href{https://www.sciencedirect.com/science/article/abs/pii/055032139390614U?via%3Dihub}{Nucl. Phys. B \textbf{399} (1993), 17-33}
\bibitem{Pepin:1996id}
S.~Pepin, F.~Stancu, M.~Genovese and J.~M.~Richard,
\href{https://www.sciencedirect.com/science/article/pii/S0370269396015973?via%3Dihub}{Phys. Lett. B \textbf{393} (1997), 119-123}
\bibitem{Janc:2004qn}
D.~Janc and M.~Rosina,
\href{https://link.springer.com/article/10.1007/s00601-004-0068-9}{Few
Body Syst. \textbf{35} (2004), 175-196}
\bibitem{Ikeda:2013vwa}
Y.~Ikeda, B.~Charron, S.~Aoki, T.~Doi, T.~Hatsuda, T.~Inoue,
N.~Ishii, K.~Murano, H.~Nemura and K.~Sasaki,
\href{https://www.sciencedirect.com/science/article/pii/S0370269314000033?via%3Dihub}{Phys. Lett. B \textbf{729}, 85-90 (2014)}
\bibitem{Carames:2011zz}
T.~F.~Carames, A.~Valcarce and J.~Vijande,
\href{https://www.sciencedirect.com/science/article/pii/S0370269311004229?via%3Dihub}{Phys. Lett. B \textbf{699}, 291-295 (2011)}
\bibitem{Molina:2010tx}
R.~Molina, T.~Branz and E.~Oset,
\href{https://journals.aps.org/prd/abstract/10.1103/PhysRevD.82.014010}{Phys.
Rev. D \textbf{82} (2010), 014010}
\bibitem{Li:2012ss}
N.~Li, Z.~F.~Sun, X.~Liu and S.~L.~Zhu,
\href{https://journals.aps.org/prd/abstract/10.1103/PhysRevD.88.114008}{Phys.
Rev. D \textbf{88} (2013) no.11, 114008}
\bibitem{Feng:2013kea}
G.~Q.~Feng, X.~H.~Guo and B.~S.~Zou,
\href{https://arxiv.org/abs/1309.7813}{[arXiv:1309.7813 [hep-ph]].}
\bibitem{Junnarkar:2018twb}
P.~Junnarkar, N.~Mathur and M.~Padmanath,
\href{https://journals.aps.org/prd/abstract/10.1103/PhysRevD.99.034507}{Phys.
Rev. D \textbf{99} (2019) no.3, 034507}
\bibitem{Maiani:2019lpu}
L.~Maiani, A.~D.~Polosa and V.~Riquer,
\href{https://journals.aps.org/prd/abstract/10.1103/PhysRevD.100.074002}{Phys.
Rev. D \textbf{100} (2019) no.7, 074002}
\bibitem{Wang:2018atz}
B.~Wang, Z.~W.~Liu and X.~Liu,
Phys. Rev. D \textbf{99}, no.3, 036007 (2019).
\bibitem{Liu:2019stu}
M.~Z.~Liu, T.~W.~Wu, M.~Pavon Valderrama, J.~J.~Xie and L.~S.~Geng,
\href{https://journals.aps.org/prd/abstract/10.1103/PhysRevD.99.094018}{Phys.
Rev. D \textbf{99} (2019) no.9, 094018}
\bibitem{Jaffe:1976yi}
R.~L.~Jaffe,
\href{https://journals.aps.org/prl/abstract/10.1103/PhysRevLett.38.195}{Phys.
Rev. Lett. \textbf{38} (1977), 195-198.}
\bibitem{Balachandran:1983dj}
A.~P.~Balachandran, A.~Barducci, F.~Lizzi, V.~G.~J.~Rodgers and
A.~Stern,
\href{https://journals.aps.org/prl/abstract/10.1103/PhysRevLett.52.887}{Phys.
Rev. Lett. \textbf{52} (1984), 887}
\bibitem{Takahashi:2001nm}
H.~Takahashi, J.~K.~Ahn, H.~Akikawa, S.~Aoki, K.~Arai, S.~Y.~Bahk,
K.~M.~Baik, B.~Bassalleck, J.~H.~Chung and M.~S.~Chung, \textit{et
al.}
\href{https://journals.aps.org/prl/abstract/10.1103/PhysRevLett.87.212502}{Phys.
Rev. Lett. \textbf{87} (2001), 212502}
\bibitem{Polinder:2007mp}
H.~Polinder, J.~Haidenbauer and U.~G.~Meissner,
\href{https://www.sciencedirect.com/science/article/pii/S0370269307009045?via%3Dihub}{Phys. Lett. B \textbf{653} (2007), 29-37}
\bibitem{Yoon:2007aq}
C.~J.~Yoon, H.~Akikawa, K.~Aoki, Y.~Fukao, H.~Funahashi, M.~Hayata,
K.~Imai, K.~Miwa, H.~Okada and N.~Saito, \textit{et al.}
\href{https://journals.aps.org/prc/abstract/10.1103/PhysRevC.75.022201}{Phys.
Rev. C \textbf{75} (2007), 022201}
\bibitem{Inoue:2010es}
T.~Inoue \textit{et al.} [HAL QCD],
\href{https://journals.aps.org/prl/abstract/10.1103/PhysRevLett.106.162002}{Phys.
Rev. Lett. \textbf{106} (2011), 162002}
\bibitem{NPLQCD:2010ocs}
S.~R.~Beane \textit{et al.} [NPLQCD],
\href{https://journals.aps.org/prl/abstract/10.1103/PhysRevLett.106.162001}{Phys.
Rev. Lett. \textbf{106} (2011), 162001}
\bibitem{Morita:2014kza}
K.~Morita, T.~Furumoto and A.~Ohnishi,
\href{https://journals.aps.org/prc/abstract/10.1103/PhysRevC.91.024916}{Phys.
Rev. C \textbf{91} (2015) no.2, 024916}
\bibitem{Li:2018tbt}
K.~W.~Li, T.~Hyodo and L.~S.~Geng,
\href{https://journals.aps.org/prc/abstract/10.1103/PhysRevC.98.065203}{Phys.
Rev. C \textbf{98} (2018) no.6, 065203}
\bibitem{Lee:2011rka}
N.~Lee, Z.~G.~Luo, X.~L.~Chen and S.~L.~Zhu,
\href{https://journals.aps.org/prd/abstract/10.1103/PhysRevD.84.014031}{Phys.
Rev. D \textbf{84} (2011), 014031}
\bibitem{Garcilazo:2020acl}
H.~Garcilazo and A.~Valcarce,
\href{https://link.springer.com/article/10.1140/epjc/s10052-020-8320-0}{Eur.
Phys. J. C \textbf{80} (2020) no.8, 720}
\bibitem{Gerasyuta:2011zx}
S.~M.~Gerasyuta and E.~E.~Matskevich,
\href{https://www.worldscientific.com/doi/abs/10.1142/S0218301312500589}{Int.
J. Mod. Phys. E \textbf{21} (2012), 1250058}
\bibitem{Froemel:2004ea}
F.~Froemel, B.~Julia-Diaz and D.~O.~Riska,
\href{https://linkinghub.elsevier.com/retrieve/pii/S0375947405000722}{Nucl.
Phys. A \textbf{750} (2005), 337-356}
\bibitem{Xia:2021tof}
Z.~Xia, S.~Fan, X.~Zhu, H.~Huang and J.~Ping,
\href{https://journals.aps.org/prc/abstract/10.1103/PhysRevC.105.025201}{Phys.
Rev. C \textbf{105} (2022) no.2, 025201}

\bibitem{Dong:2021bvy}
X.~K.~Dong, F.~K.~Guo and B.~S.~Zou,
\href{https://iopscience.iop.org/article/10.1088/1572-9494/ac27a2}{Commun.
Theor. Phys. \textbf{73} (2021) no.12, 125201}
\bibitem{Ling:2021asz}
X.~Z.~Ling, M.~Z.~Liu and L.~S.~Geng,
\href{https://link.springer.com/article/10.1140/epjc/s10052-021-09867-2}{Eur.
Phys. J. C \textbf{81} (2021) no.12, 1090}

\bibitem{Li:2012bt}
N.~Li and S.~L.~Zhu,
\href{https://journals.aps.org/prd/abstract/10.1103/PhysRevD.86.014020}{Phys.
Rev. D \textbf{86} (2012), 014020}


\bibitem{Carames:2015sya}
T.~F.~Carames and A.~Valcarce,
\href{https://journals.aps.org/prd/abstract/10.1103/PhysRevD.92.034015}{Phys.
Rev. D \textbf{92} (2015) no.3, 034015}
\bibitem{Huang:2013rla}
H.~Huang, J.~Ping and F.~Wang,
\href{https://journals.aps.org/prc/abstract/10.1103/PhysRevC.89.035201}{Phys.
Rev. C \textbf{89} (2014) no.3, 035201}
\bibitem{Meguro:2011nr}
W.~Meguro, Y.~R.~Liu and M.~Oka,
\href{https://www.sciencedirect.com/science/article/pii/S0370269311011737?via%3Dihub}{Phys. Lett. B \textbf{704}, 547-550 (2011).}




\bibitem{Wang:2021qmn}
X.~W.~Wang, Z.~G.~Wang and G.~l.~Yu,
\href{https://link.springer.com/article/10.1140/epja/s10050-021-00576-8}{Eur.
Phys. J. A \textbf{57} (2021) no.9, 275}

\bibitem{Chen:2017vai}
R.~Chen, A.~Hosaka and X.~Liu,
\href{https://journals.aps.org/prd/abstract/10.1103/PhysRevD.96.116012}{Phys.
Rev. D \textbf{96} (2017) no.11, 116012}

\bibitem{Lu:2017dvm}
J.~X.~Lu, L.~S.~Geng and M.~P.~Valderrama,
\href{https://journals.aps.org/prd/abstract/10.1103/PhysRevD.99.074026}{Phys.
Rev. D \textbf{99} (2019) no.7, 074026}

\bibitem{Bernard:1995dp}
V.~Bernard, N.~Kaiser and U.~G.~Meissner,
\href{https://www.worldscientific.com/doi/abs/10.1142/S0218301395000092}{Int.
J. Mod. Phys. E \textbf{4} (1995), 193-346}
\bibitem{Epelbaum:2008ga}
E.~Epelbaum, H.~W.~Hammer and U.~G.~Meissner,
\href{https://journals.aps.org/rmp/abstract/10.1103/RevModPhys.81.1773}{Rev.
Mod. Phys. \textbf{81} (2009), 1773-1825}
\bibitem{Machleidt:2011zz}
R.~Machleidt and D.~R.~Entem,
\href{https://www.sciencedirect.com/science/article/pii/S0370157311000457?via%3Dihub}{Phys. Rept. \textbf{503} (2011), 1-75}
\bibitem{Meissner:2015wva}
U.~G.~Mei\ss{}ner,
\href{https://iopscience.iop.org/article/10.1088/0031-8949/91/3/033005}{Phys.
Scripta \textbf{91} (2016) no.3, 033005}
\bibitem{Hammer:2019poc}
H.~W.~Hammer, S.~K\"onig and U.~van Kolck,
\href{https://journals.aps.org/rmp/abstract/10.1103/RevModPhys.92.025004}{Rev.
Mod. Phys. \textbf{92} (2020) no.2, 025004}

\bibitem{Huang:2021fdt}
B.~L.~Huang, Z.~Y.~Lin and S.~L.~Zhu,
\href{https://journals.aps.org/prd/abstract/10.1103/PhysRevD.105.036016}{Phys.
Rev. D \textbf{105} (2022) no.3, 036016}
\bibitem{Xu:2017tsr}
H.~Xu, B.~Wang, Z.~W.~Liu and X.~Liu,
\href{https://journals.aps.org/prd/abstract/10.1103/PhysRevD.99.014027}{Phys.
Rev. D \textbf{99} (2019) no.1, 014027}
\bibitem{Chen:2021cfl}
K.~Chen, R.~Chen, L.~Meng, B.~Wang and S.~L.~Zhu,
\href{https://arxiv.org/abs/2109.13057}{[arXiv:2109.13057
[hep-ph]].}



\bibitem{Chen:2021spf}
K.~Chen, B.~Wang and S.~L.~Zhu,
\href{https://arxiv.org/abs/2112.13203}{[arXiv:2112.13203
[hep-ph]].}
\bibitem{Meng:2019nzy}
L.~Meng, B.~Wang and S.~L.~Zhu,
\href{https://journals.aps.org/prc/abstract/10.1103/PhysRevC.101.064002}{Phys.
Rev. C \textbf{101}, no.6, 064002 (2020)}
\bibitem{Wang:2020dhf}
B.~Wang, L.~Meng and S.~L.~Zhu,
\href{https://journals.aps.org/prd/abstract/10.1103/PhysRevD.101.094035}{Phys.
Rev. D \textbf{101}, no.9, 094035 (2020)}
\bibitem{Scherer:2002tk}
S.~Scherer,
Adv. Nucl. Phys. \textbf{27} (2003), 277
\href{https://arxiv.org/abs/hep-ph/0210398}{[arXiv:hep-ph/0210398
[hep-ph]].}
\bibitem{Cheng:1992xi}
H.~Y.~Cheng, C.~Y.~Cheung, G.~L.~Lin, Y.~C.~Lin, T.~M.~Yan and
H.~L.~Yu,
\href{https://journals.aps.org/prd/abstract/10.1103/PhysRevD.47.1030}{Phys.
Rev. D \textbf{47}, 1030-1042 (1993).}

\bibitem{Liu:2012uw}
Z.~W.~Liu and S.~L.~Zhu,
\href{https://journals.aps.org/prd/abstract/10.1103/PhysRevD.86.034009}{Phys.
Rev. D \textbf{86}, 034009 (2012).}

\bibitem{Liu:2011xc}
Y.~R.~Liu and M.~Oka,
\href{https://journals.aps.org/prd/abstract/10.1103/PhysRevD.85.014015}{Phys.
Rev. D \textbf{85}, 014015 (2012).}
\bibitem{Meng:2018gan}
L.~Meng, G.~J.~Wang, C.~Z.~Leng, Z.~W.~Liu and S.~L.~Zhu,
\href{https://journals.aps.org/prd/abstract/10.1103/PhysRevD.98.094013}{Phys.
Rev. D \textbf{98}, no.9, 094013 (2018).}
\bibitem{Sun:2012zzd}
Z.~F.~Sun, Z.~G.~Luo, J.~He, X.~Liu and S.~L.~Zhu,
\href{https://iopscience.iop.org/article/10.1088/1674-1137/36/3/002}{Chin.
Phys. C \textbf{36} (2012), 194-204}
\bibitem{Ordonez:1995rz}
C.~Ordonez, L.~Ray and U.~van Kolck,
\href{https://journals.aps.org/prc/abstract/10.1103/PhysRevC.53.2086}{Phys.
Rev. C \textbf{53} (1996), 2086-2105}
\bibitem{Epelbaum:1999dj}
E.~Epelbaum, W.~Gloeckle and U.~G.~Meissner,
\href{https://www.sciencedirect.com/science/article/pii/S0375947499008210?via%3Dihub}{Nucl. Phys. A \textbf{671} (2000), 295-331}
\bibitem{Borasoy:1998uu}
B.~Borasoy,
\href{https://link.springer.com/article/10.1007/s100529901054}{Eur.
Phys. J. C \textbf{8} (1999), 121-130}
\bibitem{Donoghue:1998rp}
J.~F.~Donoghue and B.~R.~Holstein,
\href{https://arxiv.org/abs/hep-ph/9803312}{[arXiv:hep-ph/9803312
[hep-ph]].}
\bibitem{Donoghue:1998bs}
J.~F.~Donoghue, B.~R.~Holstein and B.~Borasoy,
\href{https://journals.aps.org/prd/abstract/10.1103/PhysRevD.59.036002}{Phys.
Rev. D \textbf{59} (1999), 036002}


\end{thebibliography}
\end{document}